\pdfoutput=1

\documentclass[11pt]{article}
\usepackage{epsf,amsmath,amssymb,graphicx,scalefnt,rotating,enumitem,fancyhdr}
\numberwithin{equation}{section}
\usepackage{acronym}

\usepackage[final]{showkeys}
\usepackage[noadjust]{cite}
\usepackage{filemod}
\usepackage{framed}
\usepackage{comment}
\usepackage[pdftex, colorlinks=true, linkcolor=black, filecolor=black,
  citecolor=black, urlcolor=black, draft=false, bookmarks,
  bookmarksnumbered=true, plainpages=false,
  linktocpage={true}]{hyperref}
\usepackage[affil-it]{authblk}
\usepackage[usenames,dvipsnames]{color}
\usepackage[capitalize]{cleveref}

\usepackage{slashed}
\usepackage{dsfont}

\newif\ifshownewacro
\includecomment{private}
\shownewacrotrue  % mark first definition of acronym in red
\newcounter{notecount}
\newcommand{\eqindentspace}{em}
\newcommand{\eqindent}[1]{\hspace{#1\eqindentspace}}

\newcommand{\trace}{\text{Tr}}
\newcommand{\chiring}{\mathring{\chi}}
\newcommand{\lmut}{L_{\mu t}}
\newcommand{\Zchiring}{\mathring{Z}_\chi}
\newcommand{\ccf}{C_\text{F}}

\newcommand{\ctr}{T_\text{R}}
\newcommand{\nc}{n_\text{c}}

\newcommand{\nf}{n_\text{f}}
\newcommand{\nh}{n_\text{h}}
\newcommand{\nl}{n_\text{l}}
\newcommand{\cca}{C_\text{A}}

\newcommand{\bare}{\text{\abbrev{B}}}
\newcommand{\citere}[1]{Ref.\,\cite{#1}}
\newcommand{\citeres}[1]{Refs.\,\cite{#1}}

\newcommand{\abbrev}[1]{{\scalefont{.9}#1}}
\newcommand{\EulerGamma}{\gamma_\text{E}}
\newcommand{\ep}{\epsilon}

\newcommand{\api}{a_\text{s}}
\newcommand{\apigf}{\hat{a}_\text{s}}
\newcommand{\dd}{\mathrm{d}}
\newcommand{\deriv}[3]{\frac{\partial\ifthenelse{\equal{#1}{}}{}{^{#1}}%
    #2}{\partial #3\ifthenelse{\equal{#1}{}}{}{^{#1}}}}
\newcommand{\dderiv}[3]{\frac{\dd\ifthenelse{\equal{#1}{}}{}{^{#1}}%
    #2}{\dd #3\ifthenelse{\equal{#1}{}}{}{^{#1}}}}
\newcommand{\order}[1]{\ensuremath{{\cal O}(#1)}}

\newcommand{\msbar}{\ensuremath{\overline{\mathrm{\abbrev{MS}}}}}

\newcommand{\myacrodef}[3]{\acrodef{#2}{#3}\newcommand{#1}{\ac{#2}}}
\myacrodef{\sftx}{SFTX}{short-flow-time expansion}
\myacrodef{\vev}{VEV}{vacuum expectation value}
\myacrodef{\rg}{RG}{renormalization group}
\myacrodef{\gff}{GFF}{gradient-flow formalism}
\myacrodef{\ope}{OPE}{operator product expansion}
\myacrodef{\qcd}{QCD}{quantum chromodynamics}
\myacrodef{\QED}{QED}{quantum electrodynamics}
\acused{QED} % do not expand this acronym in the main text
\acused{QCD} % do not expand this acronym in the main text
\myacrodef{\lhc}{LHC}{Large Hadron Collider}
\myacrodef{\uv}{UV}{ultra-violet}
\myacrodef{\lo}{LO}{leading order}
\myacrodef{\nlo}{NLO}{next-to-leading order}
\myacrodef{\nnlo}{NNLO}{next-to-next-to-leading order}
\myacrodef{\llog}{LL}{leading logarithmic}
\myacrodef{\nll}{NLL}{next-to-leading logarithmic}
\myacrodef{\nnll}{NNLL}{next-to-next-to-leading logarithmic}
\myacrodef{\pdf}{PDF}{parton density function}
\myacrodef{\sm}{SM}{Standard Model}
\myacrodef{\bsm}{BSM}{beyond-the-\ac{SM}}
\myacrodef{\mssm}{MSSM}{Minimal Supersymmetric \ac{SM}}
\myacrodef{\susy}{SUSY}{Supersymmetry}
\myacrodef{\dreg}{DREG}{Dimensional Regularization}
\myacrodef{\dred}{DRED}{Dimensional Reduction}
\allowdisplaybreaks
\newcommand{\RHheaderline}{\textsf{TTK-23-28, P3H-23-075, ZU-TH 72/23, PSI-PR-23-40, TTP23-053}
}
\fancypagestyle{firstpage}
{
  
  \fancyhead[R]{\RHheaderline}
}
\title{Short-flow-time expansion of quark bilinears through
  next-to-next-to-leading order QCD}
\author[1]{Janosch Borgulat}
\author[1]{Robert V. Harlander}
\author[1]{Jonas T. Kohnen}
\author[2,3,4,5]{Fabian~Lange}
\affil[1]{TTK, RWTH Aachen University, Sommerfeldstra\ss{}e 16, 52056 Aachen, Germany}
\affil[2]{Physik-Institut, Universit\"at Z\"urich, Winterthurerstrasse 190, 8057 Z\"urich, Switzerland}
\affil[3]{Paul Scherrer Institut, 5232 Villigen PSI, Switzerland}
\affil[4]{Institut f\"ur Theoretische Teilchenphysik, Karlsruhe Institute of Technology (KIT), Wolfgang-Gaede-Stra\ss{}e 1, 76128 Karlsruhe, Germany}
\affil[5]{Institut f\"ur Astroteilchenphysik, Karlsruhe Institute of Technology (KIT), Hermann-von-Helmholtz-Platz 1, 76344 Eggenstein-Leopoldshafen, Germany}

\date{}
\begin{document}
\maketitle
\thispagestyle{firstpage}
\begin{abstract}
  The gradient-flow formalism proves to be a useful tool in lattice
  calculations of quantum chromodynamics.  For example, it can be used as a
  scheme to renormalize composite operators by inverting the short-flow-time
  expansion of the corresponding flowed operators. In this paper, we consider
  the short-flow-time expansion of five quark bilinear operators, the scalar,
  pseudoscalar, vector, axialvector, and tensor currents, and compute the
  matching coefficients through next-to-next-to-leading order \qcd.  Among other applications, our
  results constitute one ingredient for calculating bag parameters of
  mesons within the gradient-flow formalism on the lattice.
\end{abstract}
\parskip.0cm
\tableofcontents
\parskip.2cm

%- }}}
%- {{{ body:

%\input{about-git}

%- {{{ intro:

\section{Introduction}\label{sec:intro}

The \gff\,\cite{Narayanan:2006rf,Luscher:2009eq,Luscher:2010iy} extends the
fields of \qcd{} in terms of the flow time $t$ and is meanwhile an established
tool in lattice gauge theory calculations. Its main application up to now has
been in the scale setting procedure, required to determine the lattice spacing
in physical units\,\cite{Luscher:2010iy,BMW:2012hcm}, as a scheme for defining
the strong coupling
constant\,\cite{Luscher:2010iy,Fodor:2012td,Fritzsch:2013je,
DallaBrida:2016kgh,DallaBrida:2017tru,Fodor:2017die,Hasenfratz:2019hpg}, or
simply as a smearing
mechanism\,\cite{Narayanan:2006rf,Luscher:2010iy}. However, it has been shown
that the \gff\ has a much larger potential. One of the key elements for this
is the \sftx, where composite operators of flowed fields are expressed in
terms of a regular \ope. By inversion of a complete basis of operators, this
lets one express an effective Lagrangian in regular \qcd\ in terms of flowed
operators and corresponding flowed Wilson
coefficients\,\cite{Luscher:2011bx,Suzuki:2013gza,
Makino:2014taa,Monahan:2015lha}. Matrix elements of the former do not require
renormalization\,\cite{Luscher:2011bx,Luscher:2013cpa,Hieda:2016xpq} and are
thus ideally suited to be computed on the lattice. The flowed Wilson
coefficients, on the other hand, can be obtained from the regular
\msbar\ results via suitable conversion factors, which can be calculated
perturbatively.  Obviously, the perturbative order of these conversion factors
has to match the one of the regular \msbar\ Wilson coefficients. This is why,
in many cases, \nlo\ results are not sufficient, but higher orders are
required.

The feasibility of the above approach was demonstrated via a flowed
formulation of the energy-momentum tensor in
\qcd\,\cite{Suzuki:2013gza,Makino:2014taa,Harlander:2018zpi} which was
subsequently used to extract thermodynamical observables from the
lattice\,\cite{Asakawa:2013laa,Taniguchi:2016ofw,Kitazawa:2016dsl,
Kitazawa:2017qab,Yanagihara:2018qqg,Iritani:2018idk,Taniguchi:2020mgg,
Shirogane:2020muc,Suzuki:2021tlr,Altenkort:2022yhb}.  In this case, the
coefficients of the regular operators are rational numbers. It was shown that
the \nnlo\ corrections to the matching matrix, which determines the conversion
to flowed operators, lead to a significant improvement in the extrapolation to
the physical limit at $t=0$\,\cite{Iritani:2018idk}.

As is well known from regular perturbation theory, every additional order
leads to an enormous increase in complexity. Fortunately, however, many of the
tools and techniques from regular perturbation theory can be adapted to higher
orders in the \gff. An outline of this strategy has been described
in \citere{Artz:2019bpr}, where a number of three-loop quantities were
evaluated at finite flow time. Using this approach allowed to extend the
\nlo\ results for the effective weak $|\Delta F| = 2$
Hamiltonian\,\cite{Suzuki:2020zue} or the magnetic dipole moment
operator\,\cite{Mereghetti:2021nkt} to the \nnlo{}
in \citere{Harlander:2022tgk} and \citere{Borgulat:2022cqe},
respectively. Similarly, the matching matrix between flowed and regular
operators and Wilson coefficients was obtained to the same order for the
hadronic vacuum polarization\,\cite{Harlander:2020duo}.

One of the main benefits of the \gff\ is the exponential suppression of
high-momentum modes. As already mentioned above, this implies that composite
operators of flowed fields do not require renormalization. Matrix elements of
operators which only involve flowed gluons are even finite after
renormalization of the regular \qcd\ parameters (strong coupling and
masses). Flowed quark fields, on the other hand, still require multiplicative
renormalization, typically denoted by $Z_\chi$, see \cref{sec:flow}. In order
to match perturbative and lattice results, one needs to define a suitable
renormalization scheme, most conveniently via a Green's function which
involves two flowed quark fields. One option is the so-called ringed scheme,
originally proposed in \citere{Makino:2014taa}, which fixes $Z_\chi$ via the
tree-level vacuum expectation value of the quark kinetic operator.  The
conversion factor between the \msbar{} and the ringed scheme is known through
\nnlo{}\,\cite{Artz:2019bpr}.

However, other options for the scheme of $Z_\chi$ may be more convenient. For
example, one may fix it via the \sftx\ of some quark bilinear operator,
usually called current. A preliminary lattice study of this strategy was
recently presented in \citere{Black:2023vju}, where the flowed four-quark
operator was normalized to the flowed axialvector current. Which current is most
suitable may depend on the specific calculation or observable under
consideration. It will therefore be useful to have all the associated results
at disposal.

Moreover, the simplicity of the quark currents could also be used for
systematic studies of the \sftx{}.  First, one can compare perturbative and
non-perturbative determinations of the matching coefficients with each
other. Some preliminary studies in this direction have already been carried
out in \citere{Kim:2021qae} for the \abbrev{CP}-violating quark chromoelectric
dipole moment operator and for the currents in \citere{Hasenfratz:2022wll}.
Secondly, one can compare results for the renormalized currents obtained
through the \sftx{} with results obtained in more conventional
non-perturbative schemes.  This may allow one to test non-perturbatively the
accuracy of the \sftx{} and assess the systematics associated with the $t \to
0$ limit.  A first study of higher-power terms has been done in the context of
the energy-momentum tensor in \citere{Suzuki:2021tlr}.  Besides these indirect
applications, the \sftx{} of the currents directly contribute to a number of
observables in the \gff{} such as the chiral
condensate\,\cite{Taniguchi:2016ofw} or semileptonic contributions to the
neutron electric dipole moment\,\cite{Buhler:2023gsg}.

Through \nlo{}, the \sftx\ of the currents has been calculated already several
years ago\,\cite{Endo:2015iea,Hieda:2016lly}.\footnote{While the scalar,
pseudoscalar, vector, and axialvector currents are renormalized and discussed
in more detail, for the tensor current only the bare result is provided in
\citere{Hieda:2016lly}.} In order to be consistent with the uncertainties
expected from the associated lattice calculations, one can expect that higher
orders of the matching coefficients will be relevant.  In this paper, we will
therefore derive the corresponding \nnlo{} results.

The remainder of this paper is structured as follows: In \cref{sec:framework},
we discuss the theoretical basis of our calculation, starting with the
\gff\ in \cref{sec:flow}, the definition of the regular and flowed currents in
\cref{sec:currents}, and the methods to obtain the \sftx\ in \cref{sec:sftx}.
Our results for the matching coefficients are presented in \cref{sec:results}.
The latter also allow us to evaluate the so-called flowed anomalous
dimensions, describing the logarithmic flow-time evolution of the
currents. This is presented in \cref{sec:anom}. \cref{sec:conclusions}
contains our conclusions.

%- }}}
%- {{{ section{Theoretical framework}

\section{Theoretical framework}\label{sec:framework}

%- {{{ subsection{The \qcd\ gradient flow}

\subsection{The \qcd\ gradient flow}
\label{sec:flow}

In this paper, we work in $D$-dimensional Euclidean space-time with
$D=4-2\ep$.  The \gff\ continues the gluon and quark fields $A_\mu$ and $\psi$
of regular\footnote{We use the terms ``flowed'' and ``regular'' \qcd\ to
distinguish quantities defined at $t>0$ from those defined at $t=0$. The
dependence on the $D$-dimensional space-time variable $x$ is
suppressed. $\mu,\nu,\rho,\ldots$ denote $D$-dimensional Lorentz indices,
while color and spinor indices are suppressed.}  \qcd\ to fields $B_\mu(t)$
and $\chi(t)$ through the initial conditions
\begin{equation}
  \begin{split}
    B_\mu (t=0) = A_\mu\,,\qquad \chi (t=0)=
    \psi\,,
    \label{eq:bound}
  \end{split}
\end{equation}
and the flow equations\,\cite{Narayanan:2006rf,Luscher:2010iy,Luscher:2013cpa}
\begin{equation}
  \begin{split}
    \partial_t B_\mu &= \mathcal{D}_\nu G_{\nu\mu} + \kappa
    \mathcal{D}_\mu \partial_\nu B_\nu\,,\\ \partial_t \chi &= \Delta \chi
    - \kappa\,g_\bare \partial_\mu B_\mu \chi\,,\\ \partial_t \bar
    \chi &= \bar \chi \overleftarrow \Delta + \kappa\,g_\bare \bar \chi
    \partial_\mu B_\mu\,,
    \label{eq:flow}
  \end{split}
\end{equation}
where the ``flow time'' $t$ is a parameter of mass dimension $[t]=-2$,
$g_\bare$ is the bare strong coupling, and $\kappa$ is a gauge parameter which
drops out of physical observables. In our calculations, we set $\kappa=1$.

The flowed field-strength tensor is defined as
\begin{align}
  G_{\mu\nu} = \partial_\mu B_\nu - \partial_\nu B_\mu +g_\bare [B_\mu, B_\nu]\,,
\end{align}
the flowed covariant derivative in the adjoint representation is given by
\begin{align}
  \mathcal{D}_\mu = \partial_\mu + g_\bare [B_\mu,\cdot\,] \,,
\end{align}
and
\begin{equation}
  \begin{split}
    \Delta = \mathcal{D}^\mathrm{F}_\mu \mathcal{D}^\mathrm{F}_\mu
    \,,\qquad
    \overleftarrow{\Delta} = \overleftarrow{\mathcal{D}}^\mathrm{F}_\mu
    \overleftarrow{\mathcal{D}}^\mathrm{F}_\mu \,,
    %% \label{eq:}
  \end{split}
\end{equation}
with the flowed covariant derivative in the fundamental representation,
\begin{equation}
  \mathcal{D}^\mathrm{F}_\mu = \partial_\mu + g_\bare B_\mu \,,\qquad
  \overleftarrow{\mathcal{D}}^\mathrm{F}_\mu =
    \overleftarrow{\partial}_\mu - g_\bare B_\mu \,.
\end{equation}
The flow equations can be solved perturbatively, leading to generalized
\qcd\ Feynman rules which involve exponential factors for the quark and gluon
propagators, plus additional ``flow-lines'' representing the evolution of the
fields in the flow time. The latter couple to the quarks and gluons via
``flowed vertices''.  The general formalism has been worked out
in \citeres{Luscher:2011bx,Luscher:2013cpa}, and more details can be found
in \citere{Artz:2019bpr}.

Since the flow time acts as regulator for ultraviolet divergences, the
\gff{} improves the renormalization properties.  After renormalization of the
fundamental parameters of \qcd{}, the flowed gluon field $B_\mu(t)$ is finite
and does not require field
renormalization\,\cite{Luscher:2010iy,Luscher:2011bx,Hieda:2016xpq}.  The
flowed quark fields $\chi(t)$, on the other hand, require multiplicative field
renormalization\,\cite{Luscher:2013cpa}. Throughout this paper, we will adopt
the ringed scheme\,\cite{Makino:2014taa}, where
\begin{equation}
  \label{eq::zchi-def}
  \chiring = \Zchiring^{1/2} \chi\,,\qquad \Zchiring =
  \zeta_\chi\,Z^{\msbar}_\chi\,.
\end{equation}
Both the \msbar\ expression $Z^{\msbar}_\chi$ as well as the finite conversion
factor to the ringed scheme, $\zeta_\chi$, are available through
\nnlo{}\,\cite{Luscher:2013cpa,Makino:2014taa,Harlander:2018zpi,Artz:2019bpr}.
Explicit expressions are collected in \cref{sec:renormalization}.

Parameter and field renormalization are sufficient to render physical matrix
elements of composite operators finite\,\cite{Luscher:2011bx,Luscher:2013cpa,Hieda:2016xpq}.  Thus, composite flowed operators do
not mix under
renormalization, which
enormously facilitates the lattice evaluation of their matrix elements
compared to those of regular operators, in particular if the latter mix with
operators of different mass dimension. Connection of the flowed matrix
elements to physics at $t=0$ can be made through a perturbative calculation, as will be
explained in more detail in \cref{sec:sftx}.

%- }}}
%- {{{ subsection{Quark currents}

\subsection{Quark currents}
\label{sec:currents}

In this paper, we consider $\nf=\nl+\nh$ quark flavors, where $\nl$ is the
number of massless quarks, while the remaining $\nh$ quarks have identical
mass $m$.  The bare non-diagonal and diagonal currents are defined as
\begin{equation}\label{eq::heat}
  \begin{aligned} j_{pq}^{\bare} &= \bar\psi_p\Gamma\psi_q \,,&&&
    j_{p}^{\bare} &= \bar\psi_p\Gamma\psi_p\,,
    \end{aligned}
\end{equation}
respectively, where $p,q\in\{1,\ldots,\nf\}$ are flavor
indices\footnote{Throughout the paper, sums over these flavor indices will be
explicitly indicated by the $\sum$ symbol.} with $p\neq q$. We furthermore
define the bare and renormalized singlet and non-singlet currents as
\begin{equation}\label{eq::irbm}
  \begin{aligned}
    j^{a,\text{ns}}_\bare &=
    \sum_{p\neq q}h^a_{pq}\,j^\bare_{pq}
    + \sum_{p}h^a_{pp}\,j^\bare_{p}
    \,,&&&
    j^{\text{s}}_\bare &= \sum_{p}j^\bare_{p}\,, \\
    j^{a,\text{ns}} &= Z^\text{ns} j^{a,\text{ns}}_\bare
    \,,&&&
    j^{\text{s}} &= Z^\text{s} j^{\text{s}}_\bare - 4Z_m^{-3}Z_\mathds{1}
    m_\bare^3
    \mathds{1}\,,
  \end{aligned}
\end{equation}
where $h^a$ is a traceless flavor generator, and $m_\bare$ is the bare quark
mass which is related to the $\msbar$ renormalized mass $m$ through
\begin{equation}\label{eq:framework:r}
  \begin{aligned}
    m_\bare = Z_m\,m\,,
  \end{aligned}
\end{equation}
with $Z_m\equiv Z_m^{\msbar}$ given in \cref{eq:ren:iowa,eq:ren:jeer}.
$Z^\text{s}$, $Z^\text{ns}$, and $Z_\mathds{1}$ are renormalization constants
which will be specified below. They depend on the Dirac structure
$\Gamma\in\{\Gamma_\text{S},\Gamma_\text{V}^\mu,\Gamma_\text{T}^{\mu\nu},
\Gamma_\text{A}^\mu,\Gamma_\text{P}\}$ of the current, where
\begin{equation}\label{eq::funk}
  \begin{aligned} \Gamma_\text{S} &= 1\,,&&&\qquad \Gamma_{\text V}^\mu
    &= \gamma^\mu\,,&&&\qquad \Gamma_{\text{T}}^{\mu\nu}
    &= \sigma^{\mu\nu} \equiv \frac{1}{2} \left[ \gamma^\mu,\gamma^\nu \right]\,,\\ \quad\Gamma_{\text
    A}^\mu &= \gamma^\mu\gamma_5\,,&&&\qquad \Gamma_\text{P}
    &= \gamma_5\,,
    \end{aligned}
\end{equation}
i.e.\ we consider the ``natural-parity'' scalar, vector, and tensor currents,
and the ``pseudo-parity'' axialvector and pseudoscalar currents.\footnote{This
naming is inspired by the fact that, if all Lorentz indices are chosen
spatial, the parity of the scalar, vector, and tensor currents is ``natural''
(i.e.\ $(-1)^\text{rank of tensor}$), whereas it is the opposite of that for
the axialvector and pseudoscalar currents.}  The renormalization constant
$Z_\mathds{1}$ allows for the possibility of the currents to mix with the unit
operator.

%- }}}
%- {{{ subsection{Short-flow-time expansion}\label{sec:sftx}

\subsection{Short-flow-time expansion}\label{sec:sftx}

Returning to the non-diagonal and diagonal notation, we define the flowed
currents as
\begin{equation}\label{eq::hond}
  \begin{aligned}
    \tilde{j}_{pq}(t) &= \Zchiring\,\bar\chi_p(t)\Gamma\chi_q(t)\,,
    &&&
    \tilde{j}_{p}(t) &= \Zchiring\,\bar\chi_p(t)\Gamma\chi_p(t)\,.
  \end{aligned}
\end{equation}
The flowed singlet and non-singlet currents are defined by replacing the diagonal and non-diagonal currents by their flowed versions.

The \sftx\,\cite{Luscher:2011bx} for the singlet and non-singlet current can be written as
\begin{equation}\label{eq::bleb}
  \begin{aligned}
    \tilde{j}^{a,\text{ns}}(t) &=
    \zeta_\bare^\text{ns}(t)\,j^{a,\text{ns}}_{\bare} + \mathcal{O}(t)
    \equiv
    \zeta^\text{ns}(t)\,j^{a,\text{ns}} + \mathcal{O}(t) \,,\\
    \tilde{j}^\text{s}(t) &=
    \nh m_{\bare}\left[\frac{1}{t}\zeta_\bare^{(1)}(t)
    + m^2_{\bare}\,\zeta_\bare^{(3)}(t)\right]\mathds{1}
    + \zeta_\bare^\text{s}(t)\,j^\text{s}_{\bare}+
    \mathcal{O}(t)\\
    &\equiv
    \nh m\left[\frac{1}{t}\zeta^{(1)}(t)
    + m^2\,\zeta^{(3)}(t)\right]\mathds{1}
    + \zeta^\text{s}(t)\,j^\text{s}
    + \mathcal{O}(t)\,,
  \end{aligned}
\end{equation}
where we have taken into account that we have $\nh$ massive quarks of
equal mass $m$.  The terms of order $t$ will be neglected in this paper.

The dependence of the matching coefficients $\zeta(t)$ on the flow time $t$ is
logarithmic. They are most conveniently computed by defining projectors onto
the regular-\qcd\ (i.e.\ not flowed) diagonal and non-diagonal currents. In
our case, we choose
\begin{equation}\label{eq::honk}
  \begin{aligned}
    P^{(1)}[\mathcal{O}] &= t\deriv{}{}{m_\bare}\langle
    0|\mathcal{O}|0\rangle\bigg|_{m_\bare=0}
    \,,\qquad
    P^{(3)}[\mathcal{O}] = \frac{1}{3!}\deriv{3}{}{m_\bare}\langle
    0|\mathcal{O}|0\rangle\bigg|_{m_\bare=0}\,,
  \end{aligned}
\end{equation}
and
\begin{equation}\label{eq::cnut}
  \begin{aligned}
    P_{pq}[\mathcal{O}] &=
    \mathcal{N}^{-1}\trace(\Gamma
    \mathcal{M}_{pq}(q_1,q_2))\bigg|_{q_1=q_2=m=0}\,,
  \end{aligned}
\end{equation}
where $\mathcal{M}_{pq}$ is a two-quark Green's function defined as
\begin{equation}\label{eq::howe}
  \begin{aligned}
    \int\dd^4 x\langle \psi_p(q_1)|\mathcal{O}|\psi_q(q_2)\rangle &=
    \bar{u}_p(q_1)\mathcal{M}_{pq}(q_1,q_2) u_q(q_2)\,,
  \end{aligned}
\end{equation}
the trace is over spinor and color indices, and
\begin{equation}\label{eq::holm}
  \begin{aligned}
    \mathcal{N} = \trace\left(\Gamma\Gamma\right)\,.
  \end{aligned}
\end{equation}
The nullification of the masses and external momenta in
\cref{eq::honk,eq::cnut} is understood to be taken \textit{before} any loop
integral is evaluated. This means that only tree-level diagrams contribute
when the projectors are applied to the r.h.s.\ of \cref{eq::bleb}, because all
higher-order diagrams are scaleless and thus vanish in dimensional
regularization.

\begin{figure}
  \begin{center}
    \begin{tabular}{cccc}
      \raisebox{0em}{%
        \mbox{%
          \includegraphics[%
            viewport=130 50 430 400,
            clip,width=.2\textwidth]%
                          {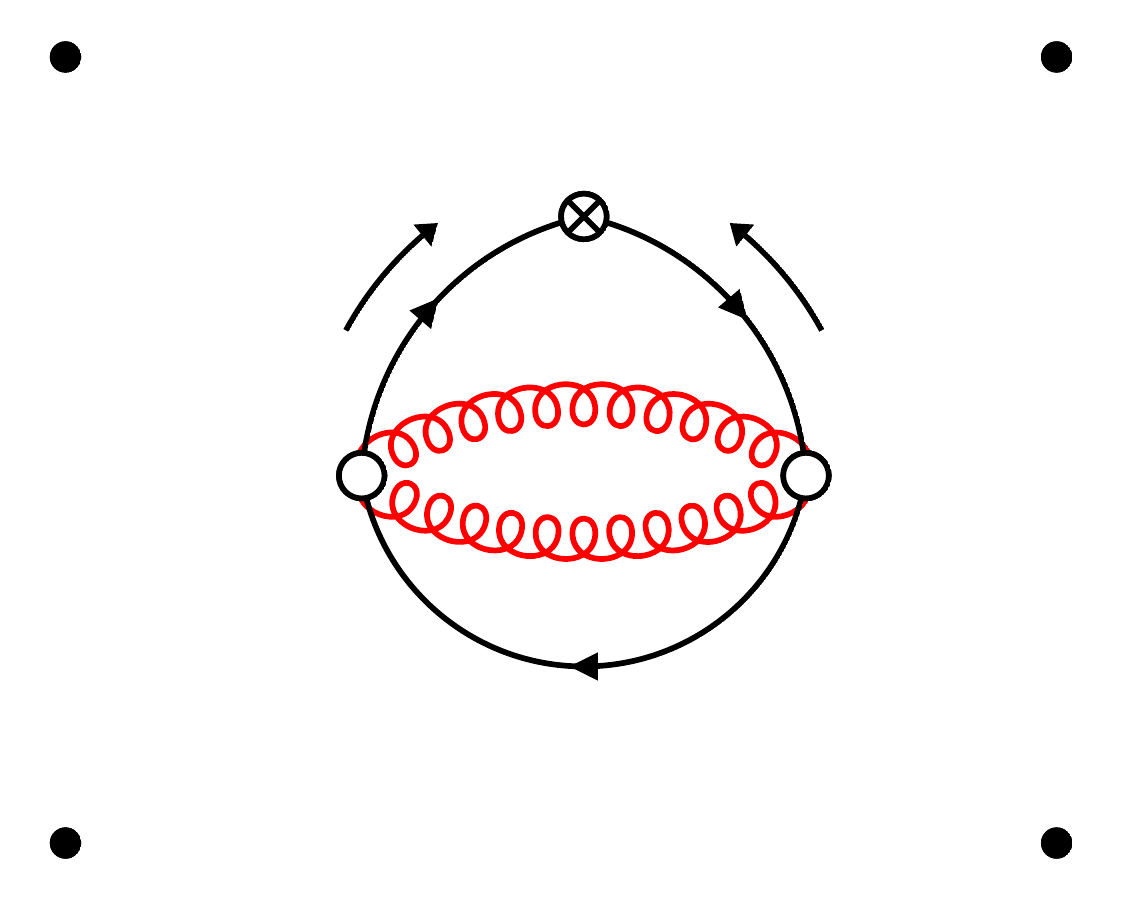}}}
      &
      \raisebox{0em}{%
        \mbox{%
          \includegraphics[%
            viewport=130 50 430 400,
            clip,width=.2\textwidth]%
                          {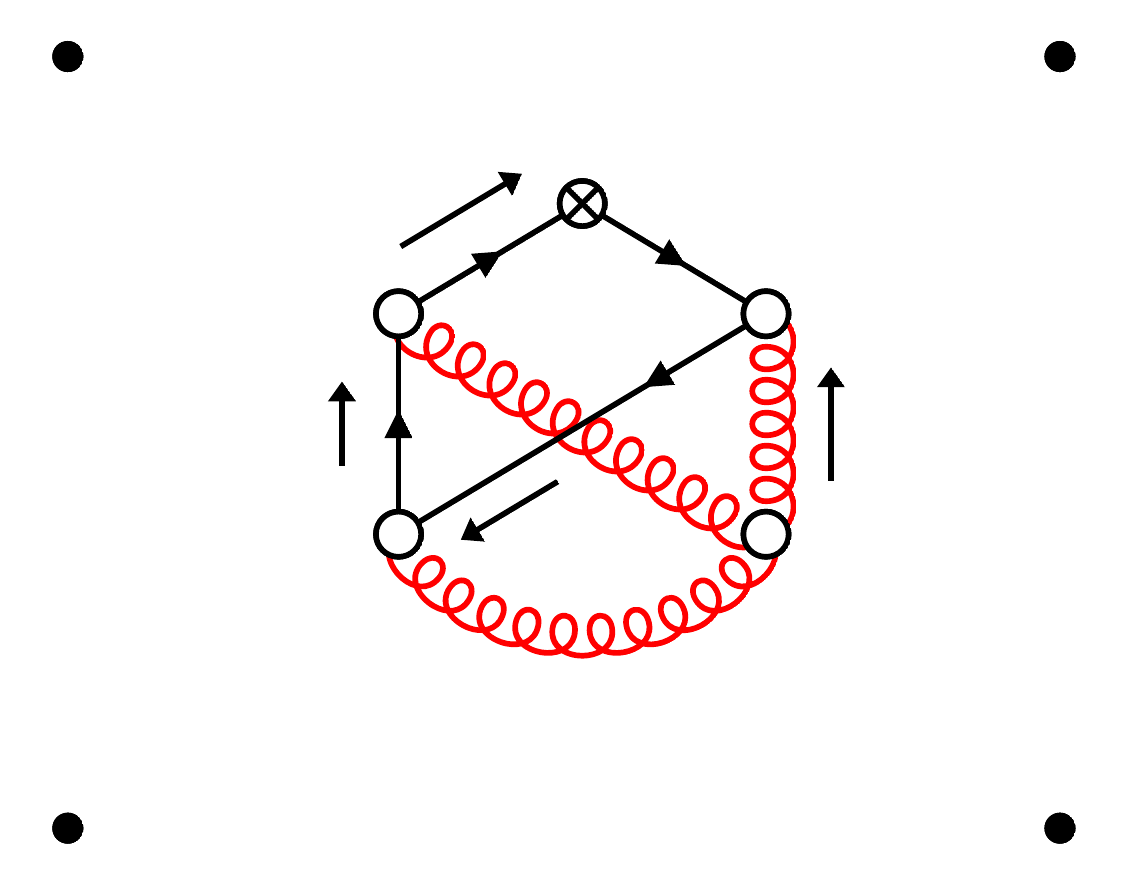}}}
      &
      \raisebox{0em}{%
        \mbox{%
          \includegraphics[%
            viewport=130 50 430 400,
            clip,width=.2\textwidth]%
                          {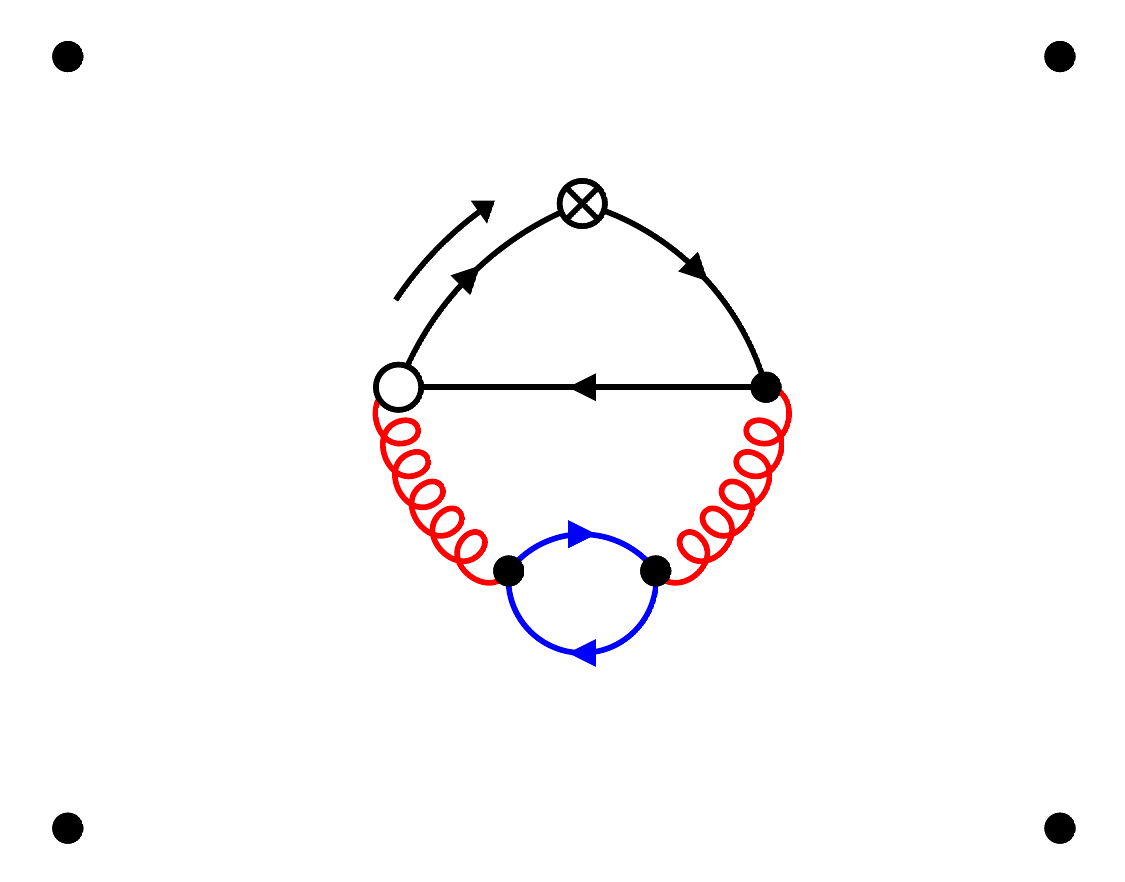}}}
      &
      \raisebox{0em}{%
        \mbox{%
          \includegraphics[%
            viewport=130 50 430 400,
            clip,width=.2\textwidth]%
                          {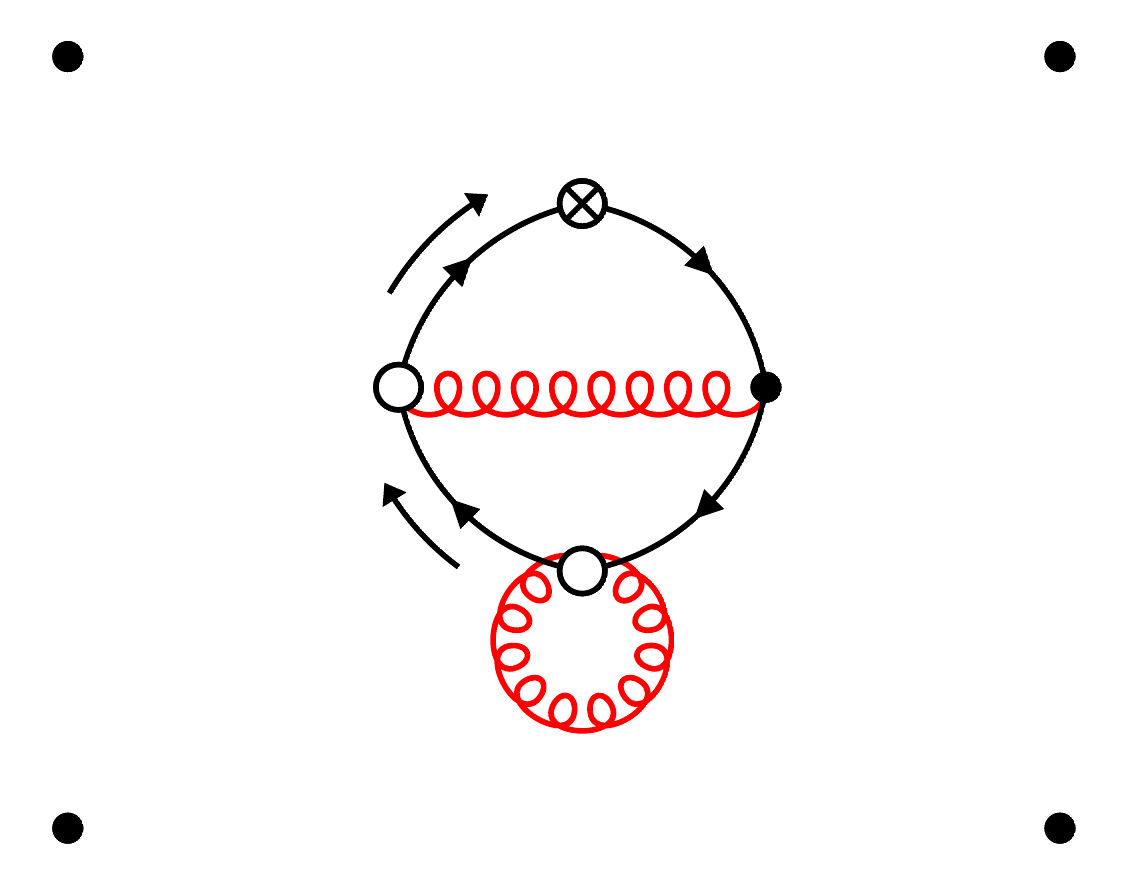}}}
      \\
      (a) & (b) & (c) & (d)
    \end{tabular}
    \parbox{.9\textwidth}{
      \caption[]{\label{fig::bubbles}\sloppy Sample diagrams contributing to
        $\zeta^{(1)}(t)$ and $\zeta^{(3)}(t)$. Spiral lines are gluons, straight
        lines denote quarks; lines with accompanying arrows are the
        corresponding flow lines (they are always connected to flowed
        vertices, denoted by small circles). The two fermion lines in
        diagram~(c) can be of different flavor. The vertex with the cross
        denotes the current. All Feynman diagrams in this paper have
        been drawn with \texttt{FeynGame}\,\cite{Harlander:2020cyh}. }}
  \end{center}
\end{figure}

The matching coefficients are then obtained as
\begin{equation}
  \begin{aligned}
  \zeta^\text{ns}_\bare(t) = P_{pq}[\tilde{j}_{pq}] \qquad \text{and} \qquad
  \zeta^\text{s}_\bare(t)  = \zeta^\text{ns}_\bare(t) + \nf \zeta^\Delta_\bare(t) \, ,
  \end{aligned}
\end{equation}
where
\begin{equation}
\zeta^\Delta_\bare(t) = P_{pp}[\tilde{j}_{q}]\,,
\end{equation}
with $p\neq q$, and
\begin{equation}\label{eq::jake}
  \begin{aligned}[t]
      \zeta^{(1)}_\bare(t) &= P^{(1)}[\tilde{j}_p]\,,
      &&&
      \zeta^{(3)}_\bare(t) &=
      P^{(3)}[\tilde{j}_p]\,,
  \end{aligned}
\end{equation}
for a massive quark flavor $p$. The renormalized matching coefficients follow
from this by inserting \cref{eq::irbm} into \cref{eq::bleb}:
\begin{equation}\label{eq::anax}
  \begin{aligned}
  \zeta^{(1)}(t) &= Z_m\,\zeta^{(1)}_\bare(t)\,,\\
  \zeta^{(3)}(t) &= Z^3_m \zeta^{(3)}_\bare(t) +\frac{4}{\nh}Z_\mathds{1}(Z^\text{s})^{-1} \zeta^\text{s}_\bare(t)
  \,,\\
    \zeta^\text{ns}(t) &= (Z^\text{ns})^{-1}\zeta_\bare^\text{ns}(t)\,,\\
    \zeta^\text{s}(t) &=
    (Z^\text{s})^{-1}
    \left[\zeta_\bare^\text{ns}(t) + \nf \zeta_\bare^{\Delta}(t)\right]
    = (Z^\text{s})^{-1}Z^\text{ns}\zeta^\text{ns}(t)
    + \nf(Z^\text{s})^{-1}\zeta_\bare^{\Delta}(t)\\& \equiv \zeta^\text{ns}(t) +
    \nf\zeta^\Delta(t)\,,\\
    \Rightarrow\ \zeta^\Delta(t) &=
    (Z^\text{s})^{-1}\left[\zeta_\bare^\Delta(t) -
      \frac{1}{\nf}(Z^\text{s}-Z^\text{ns})\zeta^\text{ns}(t)\right]\,.
  \end{aligned}
\end{equation}
$\zeta_\bare^{(1)}(t)$ and $\zeta_\bare^{(3)}(t)$ are just given by the first two
terms in an expansion in $m_\bare^2t$ of the currents' vacuum expectation
values. Due to Lorentz and parity invariance, they are non-zero only for the
scalar current, corresponding to the so-called quark condensate. Since the
one-loop contribution is of order $g_\bare^0$, it is required to three-loop
order in order to obtain the \sftx\ up to \nnlo\ \qcd. Sample three-loop
diagrams are shown in \cref{fig::bubbles}.

%- {{{ fig::bubbles

%
%

%- }}}

\begin{figure}[ht]
  \begin{center}
    \begin{tabular}{cccc}
      \raisebox{0em}{%
        \includegraphics[%
          viewport=100 50 400 400,
          clip,width=.2\textwidth]%
        {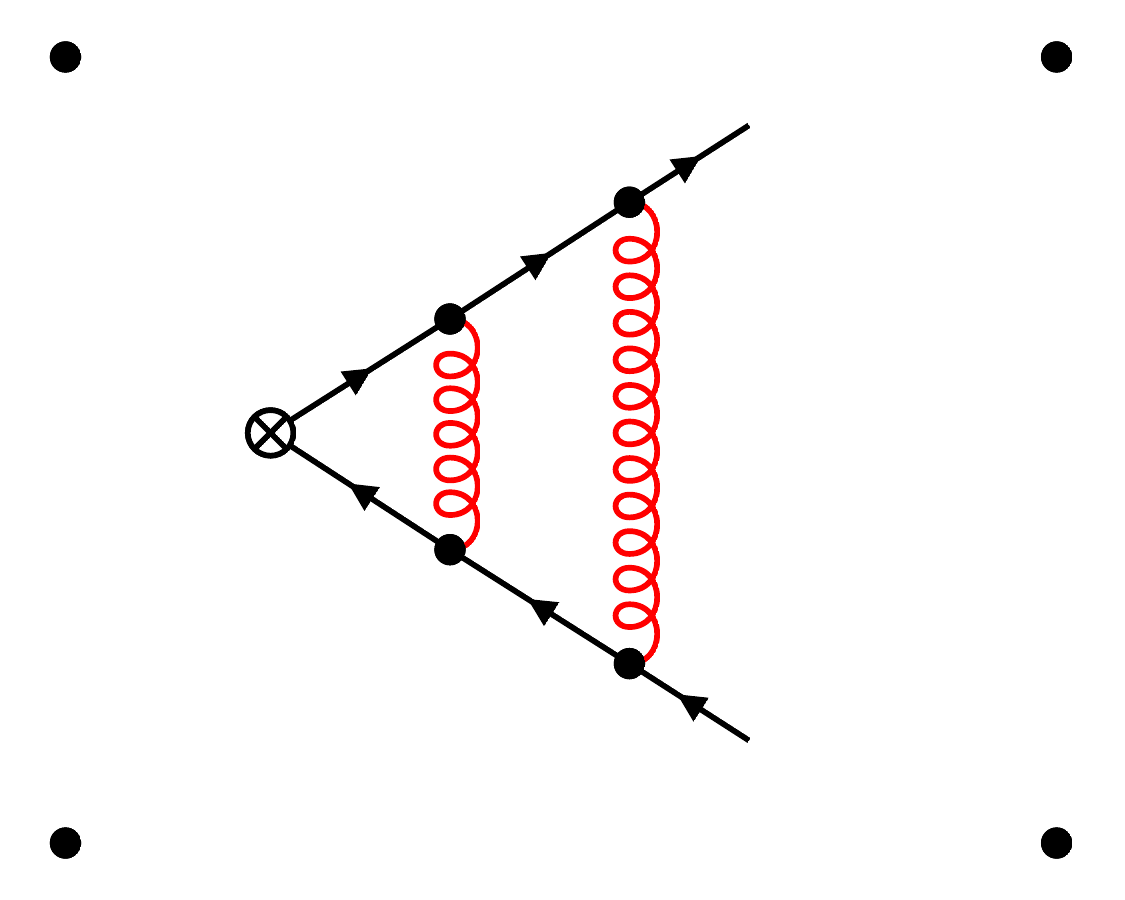}}
      &
      \raisebox{0em}{%
        \includegraphics[%
          viewport=100 50 400 400,
          clip,width=.2\textwidth]%
                        {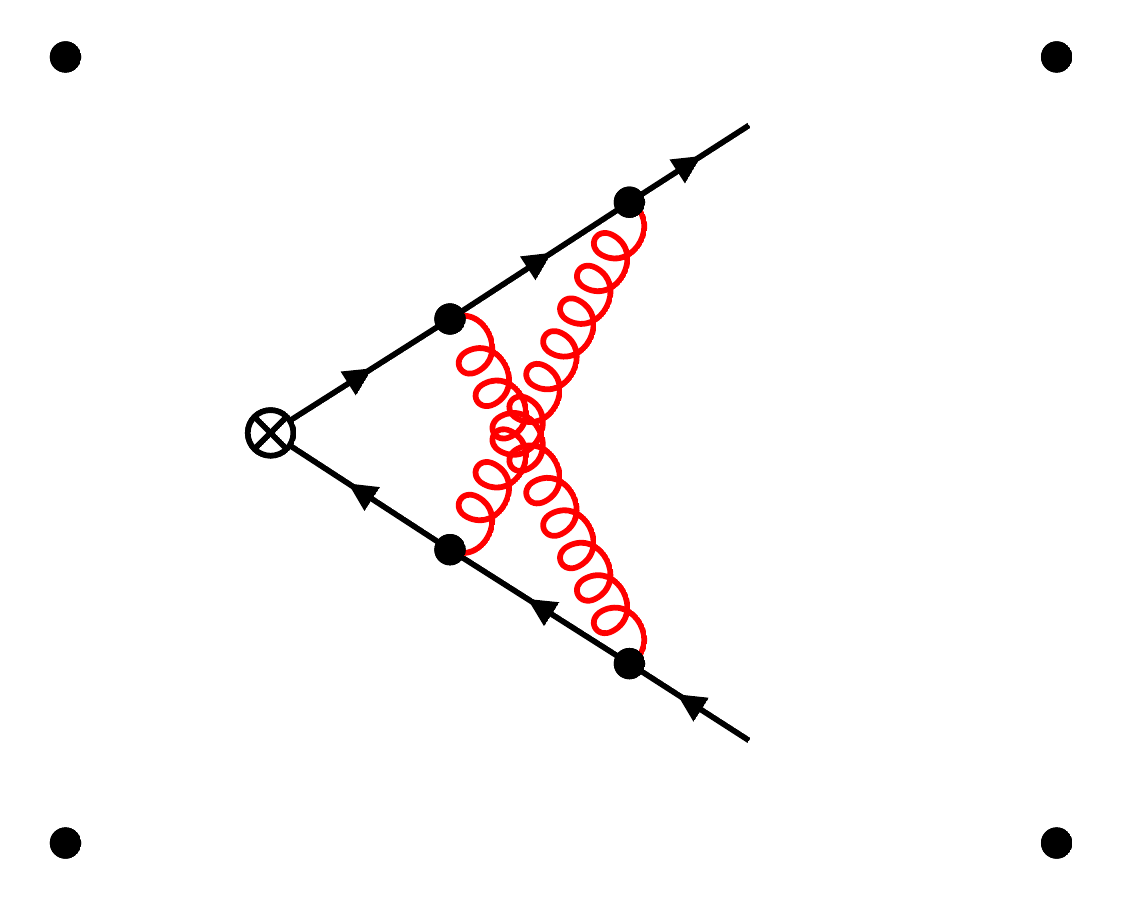}}
      &
      \raisebox{0em}{%
        \includegraphics[%
          viewport=100 50 400 400,
          clip,width=.2\textwidth]%
                        {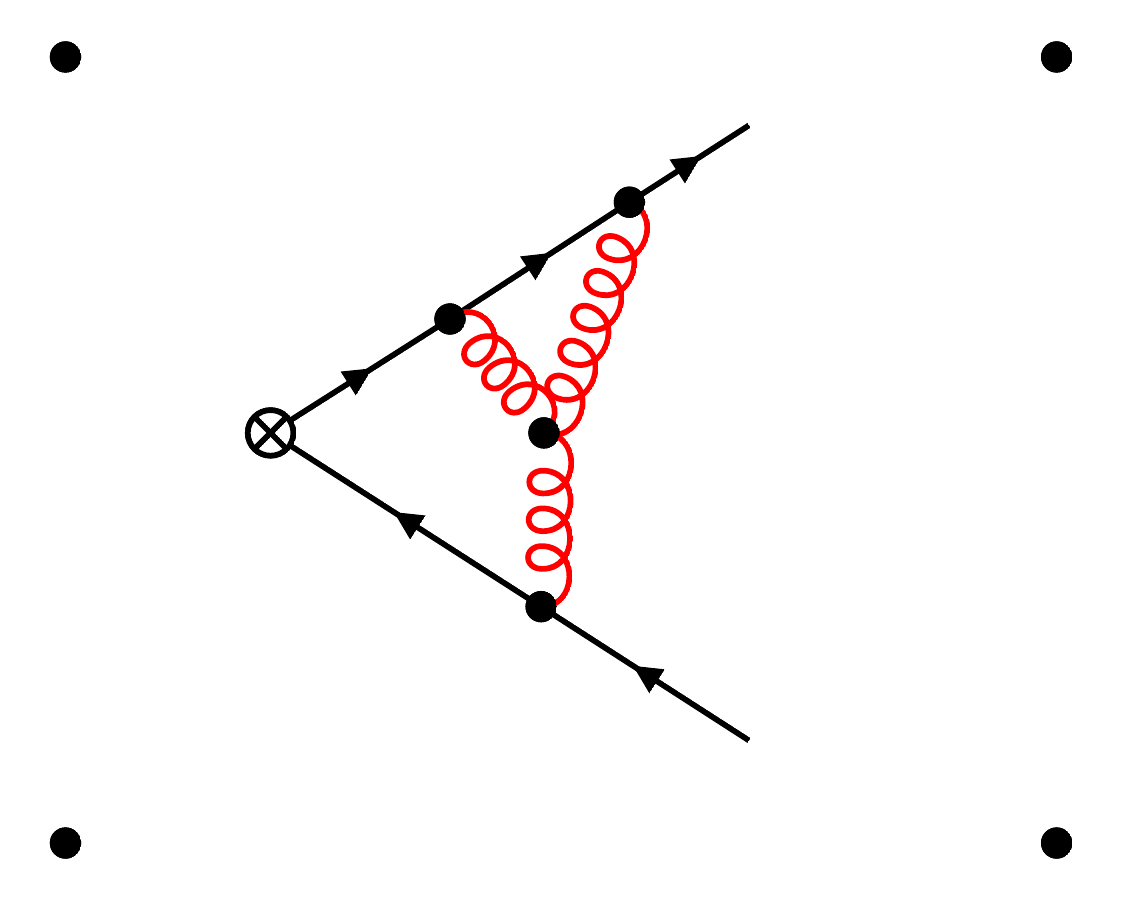}}
      &
      \raisebox{0em}{%
        \includegraphics[%
          viewport=100 50 400 400,
          clip,width=.2\textwidth]%
                        {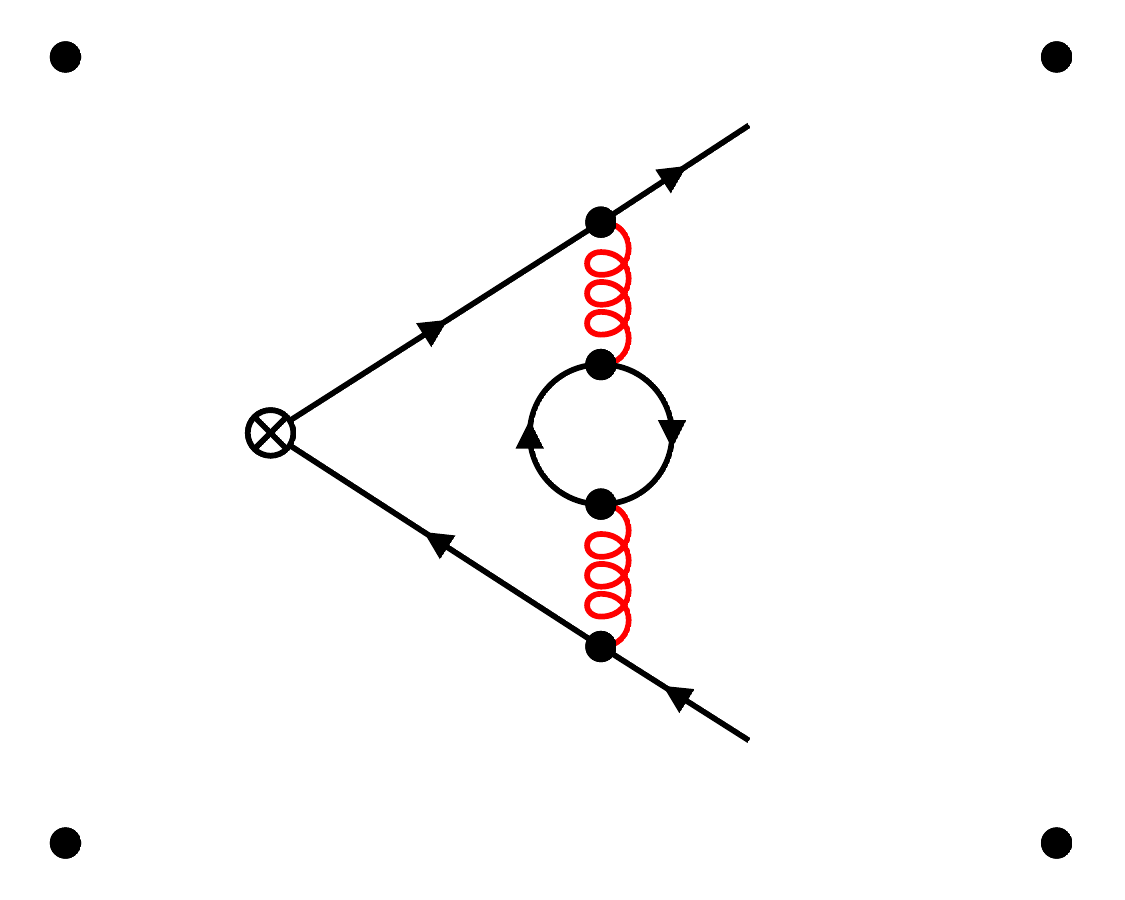}}
      \\
      \raisebox{0em}{%
        \includegraphics[%
          viewport=100 50 400 400,
          clip,width=.2\textwidth]%
        {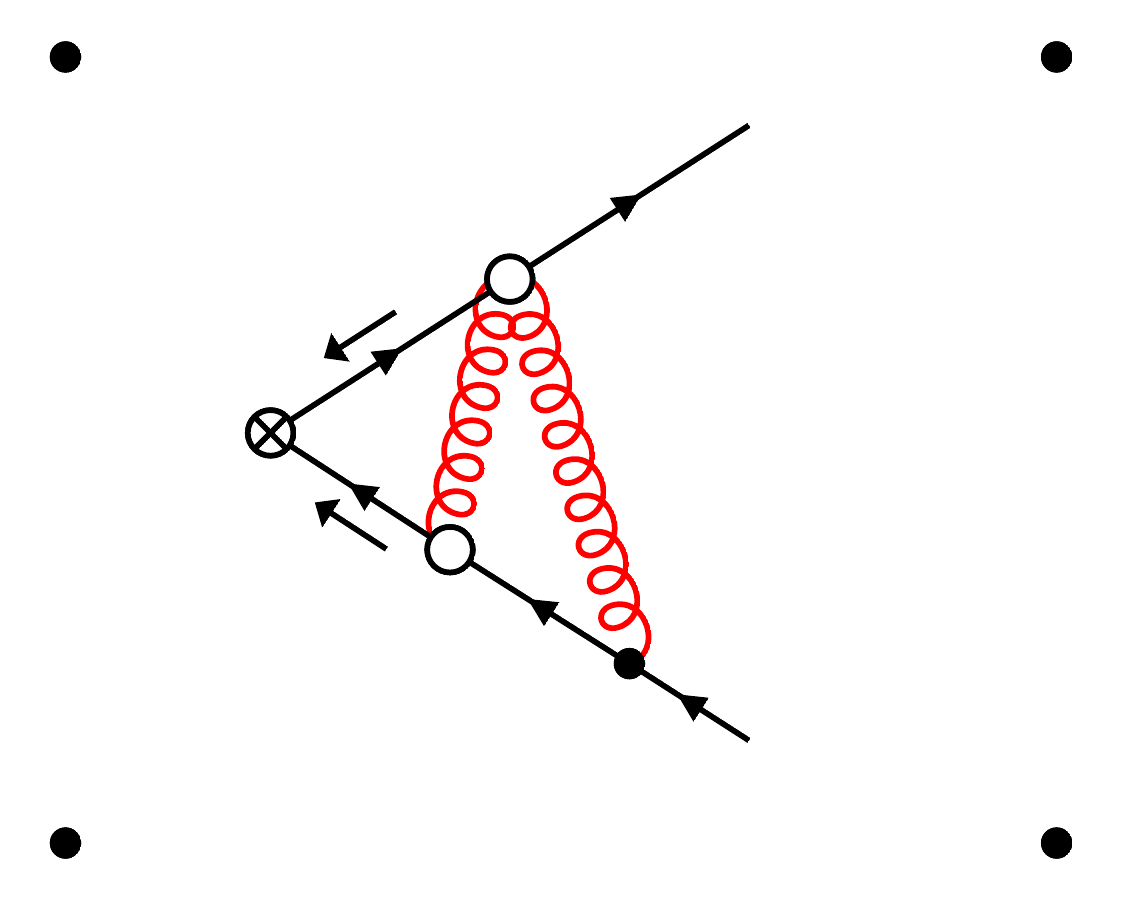}}
      &
      \raisebox{0em}{%
        \includegraphics[%
          viewport=100 50 400 400,
          clip,width=.2\textwidth]%
                        {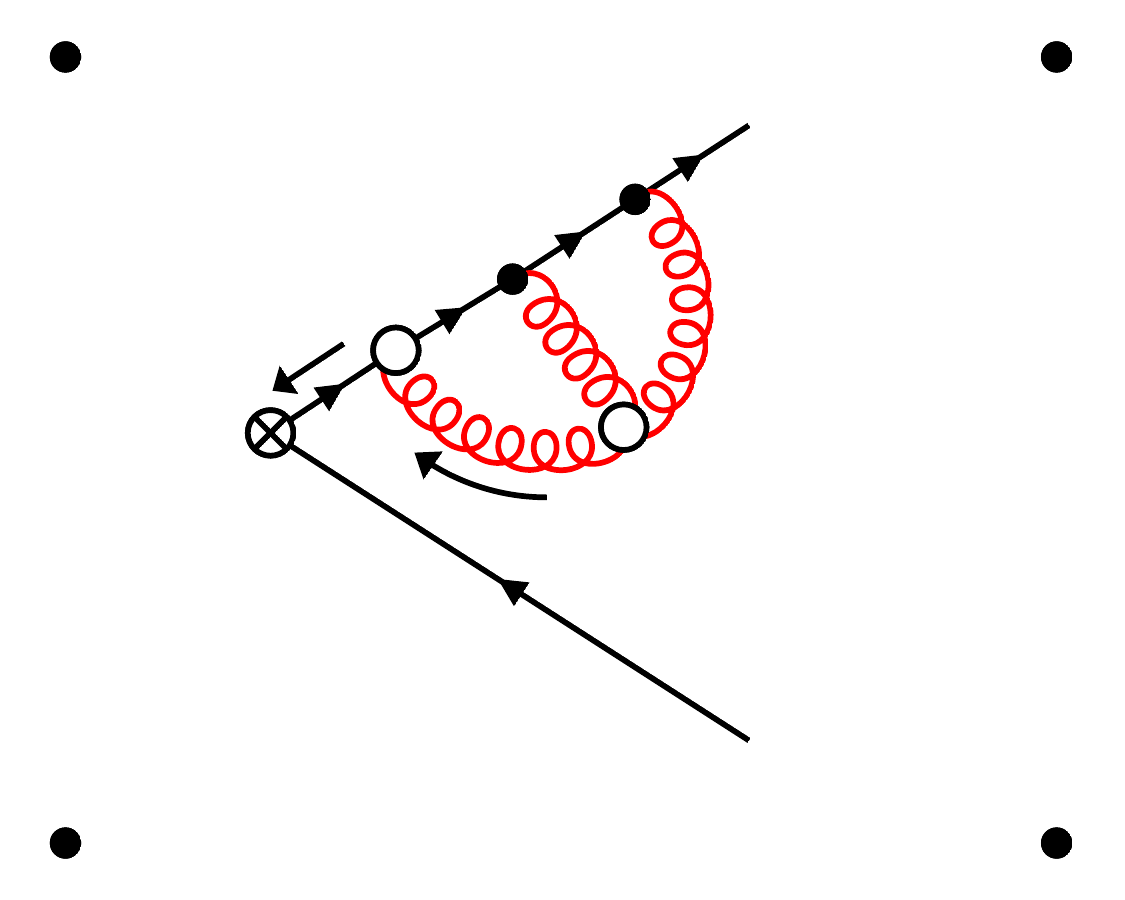}}
      &
      \raisebox{0em}{%
        \includegraphics[%
          viewport=100 50 400 400,
          clip,width=.2\textwidth]%
                        {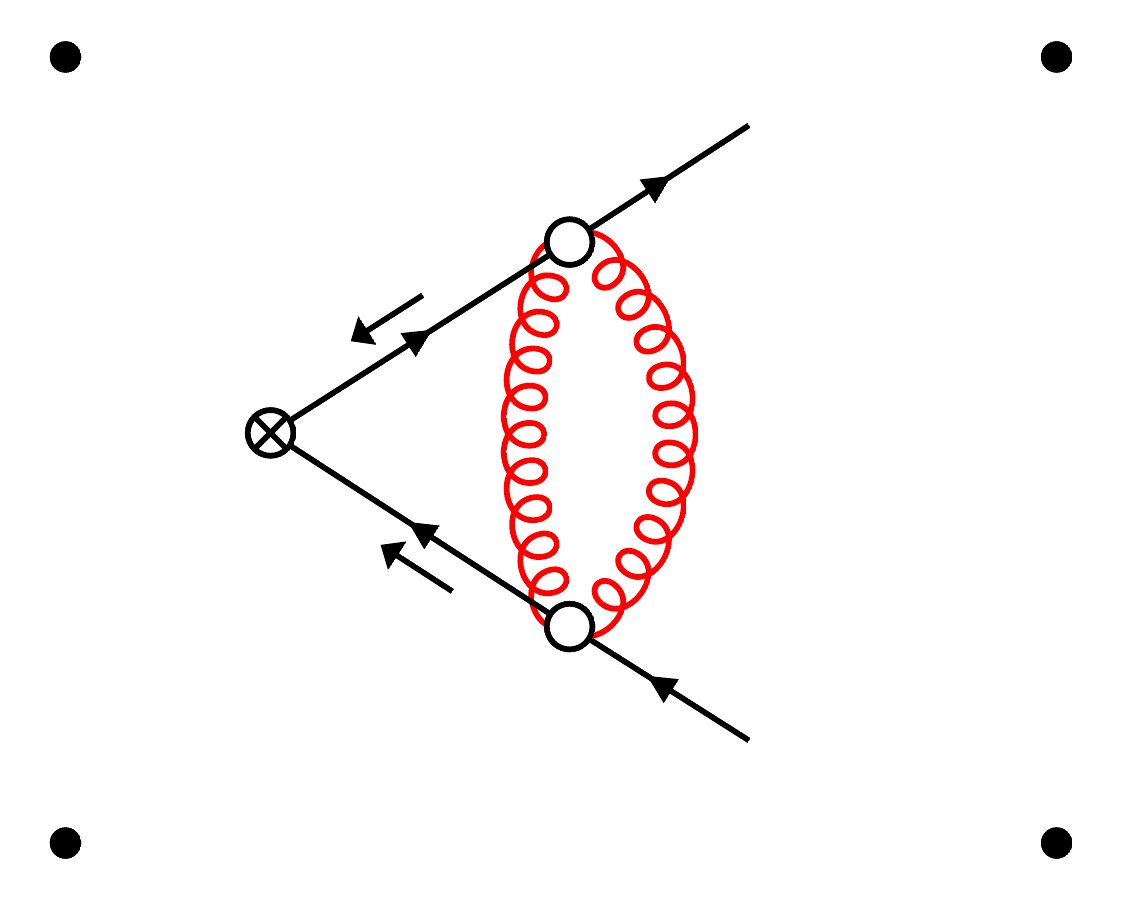}}
      &
      \raisebox{0em}{%
        \includegraphics[%
          viewport=100 50 400 400,
          clip,width=.2\textwidth]%
                        {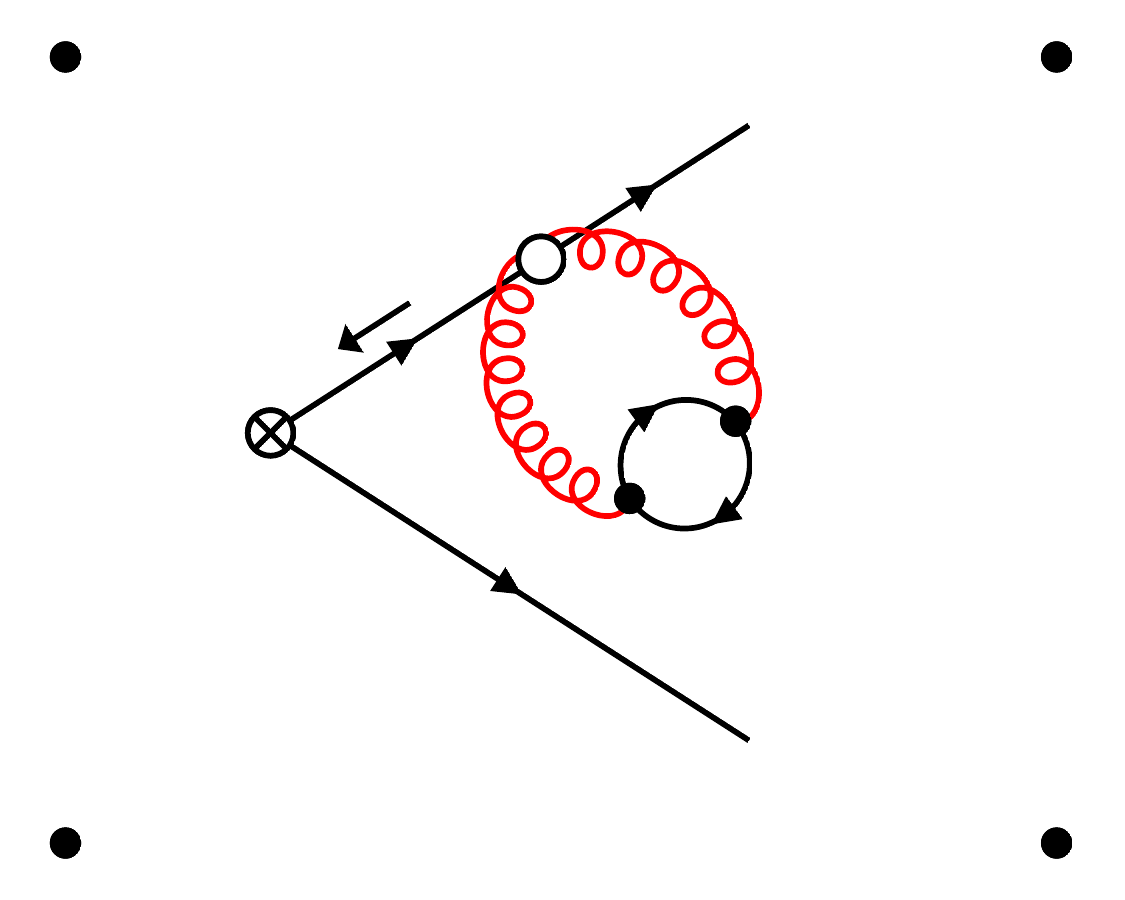}}
    \end{tabular}
    \parbox{.9\textwidth}{
      \caption[]{\label{fig::nonsing_dias}\sloppy
        Examples for contributions to the non-singlet matching
        coefficients. The notation is the same as in \cref{fig::bubbles}.
    }}
  \end{center}
\end{figure}

\begin{figure}[ht]
  \begin{center}
    \begin{tabular}{cc}
      \raisebox{0em}{%
          \includegraphics[%
            viewport=100 50 400 400,
            clip,width=.2\textwidth]%
                          {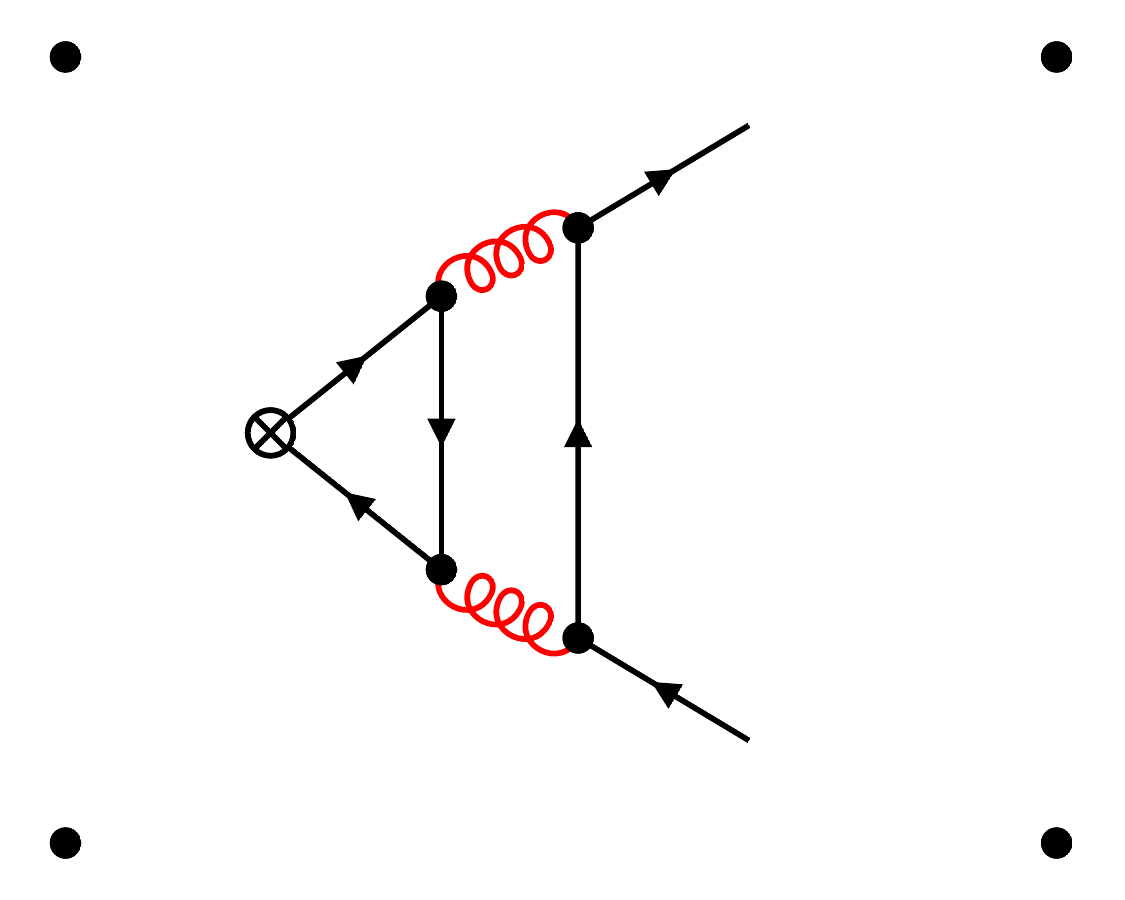}}
      &
      \raisebox{0em}{%
        \includegraphics[%
          viewport=100 50 400 400,
          clip,width=.2\textwidth]%
                        {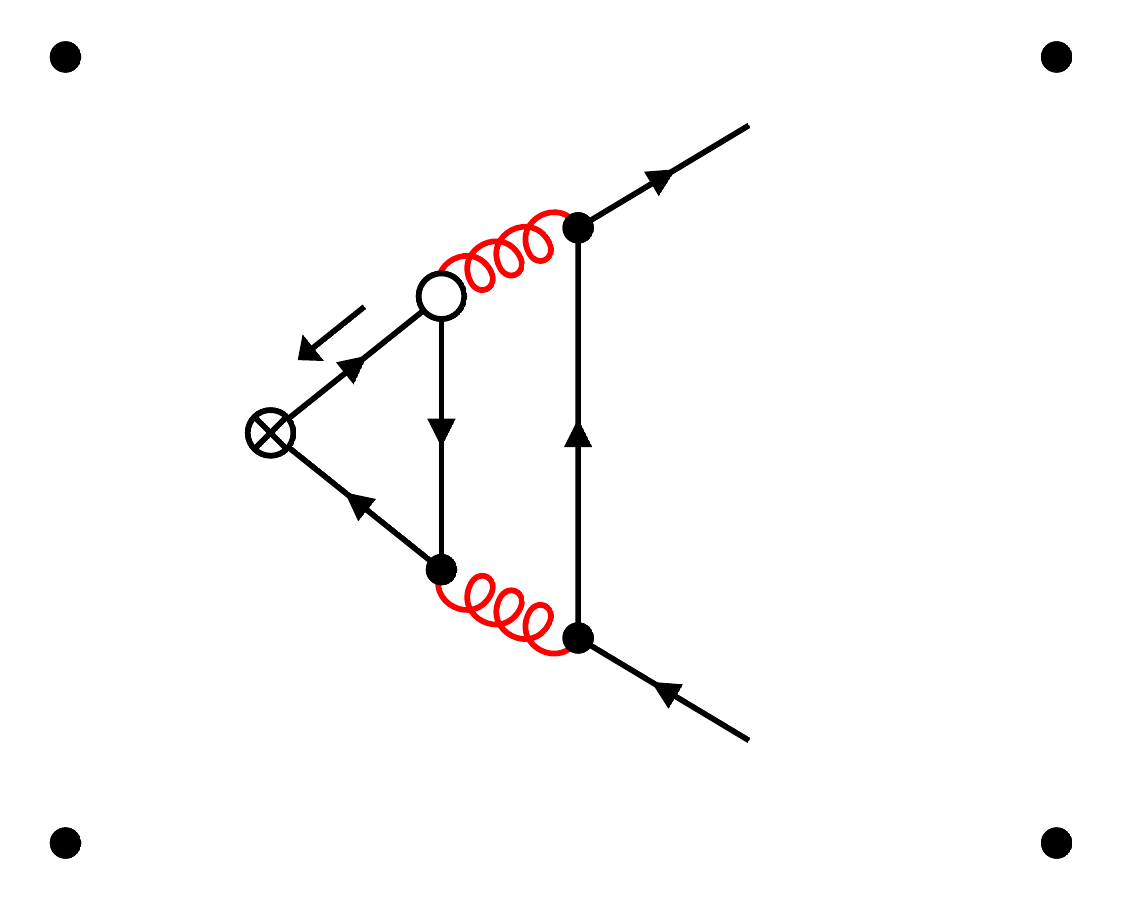}}
    \end{tabular}
    \parbox{.8\textwidth}{
      \caption[]{\label{fig::sing_dias}\sloppy Examples for contributions to
        the singlet matching coefficients.  The notation is the same as in
        \cref{fig::bubbles}.  }}
  \end{center}
\end{figure}

In contrast, the projections for the other matching coefficients in \cref{eq::jake} require only two-loop calculations.
In the diagrams that contribute to $\zeta^\text{ns}_\bare(t)$, the current is
connected to the external states by a single quark line, cf.\ \cref{fig::nonsing_dias}, as opposed to
$\zeta_\bare^{\Delta}(t)$, where the current and the external states belong to
different quark lines, cf.\ \cref{fig::sing_dias}.

%- {{{ fig::nonsing_dias:

%- }}}
%- {{{ fig::sing_dias:

%

%- }}}

%
We will refer to the latter class as \textit{triangle diagrams} in what follows.
The triangle diagrams for the scalar, pseudoscalar, and tensor currents
vanish after taking the fermion trace.  For the vector current, they only
start to contribute from three-loop order due to Furry's theorem. At the
perturbative order considered here, we can therefore drop the superscripts
``ns'' and ``s'' in these cases and simply write
\begin{equation}\label{eq::eild}
  \begin{aligned}
    \zeta_\text{X}(t)\equiv \zeta_\text{X}^\text{ns}(t) =
    \zeta^\text{s}_\text{X}(t)
    \quad \text{for}\quad \text{X}\neq \text{A}\,,
  \end{aligned}
\end{equation}
and analogously for the bare matching coefficients. For the axial current, on
the other hand, we will find $\zeta_\bare^\Delta(t)\neq 0$ at the two-loop level.

We evaluate all diagrams in $D=4-2\ep$ space-time dimensions.  The occurrence
of $\gamma_5$ in \cref{eq::funk} causes the well-known complications which we
take care of by following the strategy outlined in
\citeres{tHooft:1972tcz,Larin:1991tj,Larin:1993tq}. This means to replace
\begin{equation}\label{eq::iaso}
  \begin{aligned}
    \Gamma_\text{A}^\mu &\to \hat{\Gamma}_\text{A}^\mu =
    \frac{1}{3!}\varepsilon^{\mu\alpha\beta\gamma}
    \gamma_\alpha\gamma_\beta\gamma_\gamma\,,
    &&&
    \Gamma_\text{P} &\to \hat{\Gamma}_\text{P} =
    \frac{1}{4!}\varepsilon^{\alpha\beta\gamma\delta}
    \gamma_\alpha\gamma_\beta\gamma_\gamma\gamma_\delta
  \end{aligned}
\end{equation}
both in the currents of \cref{eq::hond} as well as in the projectors of
\cref{eq::cnut}. The resulting products of two (intrinsically four-dimensional)
$\varepsilon$ tensors are replaced by
\begin{equation}\label{eq::dreg}
  \begin{aligned}
    \varepsilon^{\alpha\beta\gamma\delta}
    \varepsilon_{\alpha'\beta'\gamma'\delta'} &=
    g^{[\alpha\phantom{]}}_{\phantom{[}\alpha'\phantom{]}}
    g^{\phantom{[}\beta\phantom{]}}_{\phantom{[}\beta'\phantom{]}}
    g^{\phantom{[}\gamma\phantom{]}}_{\phantom{[}\gamma'\phantom{]}}
    g^{\phantom{[}\delta]}_{\phantom{[}\delta'\phantom{]}}\,,&&&
    \varepsilon^{\mu\alpha\beta\gamma}
    \varepsilon_{\mu\alpha'\beta'\gamma'} &=
    g^{[\alpha\phantom{]}}_{\phantom{[}\alpha'\phantom{]}}
    g^{\phantom{[}\beta\phantom{]}}_{\phantom{[}\beta'\phantom{]}}
    g^{\phantom{[}\gamma]}_{\phantom{[}\gamma'\phantom{]}}\,,
  \end{aligned}
\end{equation}
where the square brackets denote the anti-symmetric combination, e.g.
\begin{equation}\label{eq::galt}
  \begin{aligned}
    g^{[\alpha\phantom{]}}_{\phantom{[}\alpha'\phantom{]}}
    g^{\phantom{[}\beta]}_{\phantom{[}\beta'\phantom{]}}
    = g^\alpha_{\alpha'}g^\beta_{\beta'}
    - g^\beta_{\alpha'}g^\alpha_{\beta'}\,.
  \end{aligned}
\end{equation}
This also affects the normalization factors of \cref{eq::holm}
via\footnote{Recall that the trace also includes color.}
\begin{equation}\label{eq::dos1}
  \begin{aligned}
    \trace(\hat{\Gamma}_\text{A}^\mu\hat{\Gamma}_\text{A}^\mu) &=
    -\frac{2\nc}{3}D(D-1)(D-2)\,,\\
    \trace(\hat{\Gamma}_\text{P}\hat{\Gamma}_\text{P}) &=
    -\frac{\nc}{3!}D(D-1)(D-2)(D-3)\,.
  \end{aligned}
\end{equation}
This strategy violates the Ward identities, but they can be restored by an
additional finite renormalization discussed below.

For the actual calculation, we adopt the framework developed in
\citere{Artz:2019bpr}, which is based on
\texttt{qgraf}\,\cite{Nogueira:1991ex,Nogueira:2006pq} for the generation of
the diagrams, \texttt{q2e/exp}\,\cite{Harlander:1998cmq,Seidensticker:1999bb}
for inserting the Feynman rules and identifying the momentum topologies,
in-house \texttt{FORM}\,\cite{Vermaseren:2000nd,Kuipers:2012rf,Ruijl:2017dtg}
routines for performing various computer algebraic operations including Dirac
and color algebra\,\cite{vanRitbergen:1998pn}, and
\texttt{Kira}$\otimes$\texttt{FireFly}\,\cite{Maierhofer:2017gsa,Klappert:2020nbg,
  Klappert:2019emp,Klappert:2020aqs} for the reduction to
master integrals employing integration-by-parts-like
relations\,\cite{Tkachov:1981wb,Chetyrkin:1981qh,Artz:2019bpr} and the Laporta
algorithm\,\cite{Laporta:2000dsw}. Up to two-loop level, we find the same
master integrals as in \citere{Harlander:2018zpi}. They can be evaluated
analytically in terms of the transcendentals\footnote{We caution the reader
not to confuse the multiple use of the symbol $\zeta$ in this paper.}
\begin{equation}\label{eq:framework:gene}
  \begin{aligned}
    \zeta_2=\text{Li}_2(1) &= \frac{\pi^2}{6} = 1.64493\ldots\,,\qquad
    \zeta_3 = \text{Li}_3(1) = 1.20205\ldots\,,\\
    \text{Li}_2(1/4) &= 0.267652\ldots\,,
  \end{aligned}
\end{equation}
where $\text{Li}_n(z)=\sum_{k=1}^\infty z^k/k^n$ is the polylogarithm of order
$n$.

The three-loop vacuum expectation value contributing to $\zeta^{(1)}(t)$ and
$\zeta^{(3)}(t)$ of \cref{eq::anax} leads to $304$ diagrams which are reduced to
216 master integrals. They have been evaluated numerically in \citere{Harlander:2020duo} following the strategy
described in \citere{Harlander:2016vzb}.

%- }}}

%- }}}
%- {{{ section{Results}

\section{Results}\label{sec:results}

%- {{{ subsection{Natural-parity currents}

\paragraph{Natural-parity currents.}
For the currents which do not involve $\gamma_5$, we adopt the \msbar\ scheme,
i.e., we define the renormalization constants of
\cref{eq::heat} as
\begin{equation}\label{eq::berm}
  \begin{aligned}
    Z_\text{X} &= Z_\text{X}^{\msbar}\,,\qquad
    \text{X}\in\{\text{S},\text{V},\text{T}\}\,.
  \end{aligned}
\end{equation}
Because of Lorentz and parity invariance, only the scalar current can mix with the identity operator. Thus, we have
\begin{equation}
  \begin{aligned}
    Z_{\mathds{1},\text{S}}=Z_{\mathds{1},\text{S}}^{\msbar} &=
    \left(\frac{\mu^2e^{\EulerGamma}}{4\pi}\right)^{-\ep}
    Z_0 \qquad \text{and} \qquad Z_{\mathds{1},\text{X}}=0\,,\qquad
    \text{X}\in\{\text{V},\text{P},\text{A},\text{T}\}\,,
  \end{aligned}
\end{equation}
where $Z_0$ is the renormalization constant of the vacuum energy which can be
found in \cref{eq::z0}, $\mu$ is the renormalization scale in the $\msbar$
scheme, and $\EulerGamma = -\Gamma'(1)=0.577216\ldots$ is the Euler-Mascheroni
constant, with Euler's gamma function $\Gamma(z)$. The Ward-Takahashi
identities ensure that $Z_\text{V}^{\msbar}=1$ and $Z_\text{S}^{\msbar}=Z_m$,
with $Z_m$ the quark mass renormalization constant introduced above. For the
tensor current, the renormalization constant is given in
\cref{eq::fold,eq::anom2}.

We express our results in terms of the color factors
\begin{equation}\label{eq::joss}
  \begin{aligned}
    \ccf &= \ctr\,\frac{\nc^2-1}{\nc}\,,\qquad \cca=2\ctr\nc\,,\qquad
    \ctr=\frac{1}{2}\,,
  \end{aligned}
\end{equation}
where in \qcd, the number of colors is $\nc=3$. Furthermore, we introduce
\begin{equation}\label{eq:calculation:fane}
  \begin{aligned}
    \lmut = \ln 2\mu^2 t + \EulerGamma \equiv \ln\frac{\mu^2}{\mu_t^2}\,,
  \end{aligned}
\end{equation}
where we have implicitly defined the $t$-dependent energy scale $\mu_t$, and
\begin{equation}\label{eq:results:acis}
  \begin{aligned}
    \api = \frac{g^2}{4\pi^2} = \frac{\alpha_s}{\pi}\,,
  \end{aligned}
\end{equation}
with $g$ the $\msbar$ renormalized strong coupling, see \cref{eq::gren}.  We
then find the following matching coefficients for the natural-parity currents:
\begin{align*}
\zeta_\text{S}^{(1)}(t)
  &= -\frac{\nc}{8\pi^2} \Bigg\{ 1  + \api\,\ccf\Big( 1 + \ln2
    - \frac{3}{4}\ln3
    + \frac{3}{4}\lmut\Big)\\
    &
     \eqindent{1} + \api^2\bigg[
    1.228\,\ccf^2 + 2.587\,\cca\ccf -0.9873\,\ccf\ctr\nf
    \\&\eqindent{2}
    + \lmut\big(0.7456\,\ccf^2 + 1.807\,\cca\ccf
      - 0.4981 \,\ccf\ctr\nf \big)
    \\&\eqindent{2}
    + \lmut^2\big(0.2813\,\ccf^2 + 0.3438\,\cca\ccf
      - 0.1250 \,\ccf\ctr\nf \big) \bigg] \Bigg\} + \mathcal{O}(\api^3)\,,
    \stepcounter{equation}\tag{\theequation}\label{eq::zetaS1}\\
    \zeta_\text{S}^{(3)}(t)
    &= -\frac{\nc}{4\pi^2} \Bigg\{ 1
    + \lmut
    + \api\,\ccf\Bigg[ \frac{7}{2}
    + 4 \ln2
    -\frac{21}{4}\ln3
    - 3 \text{Li}_2\left(1/4\right)
    \\
    & \eqindent{2}
    + \lmut\bigg(\frac{11}{4}-\ln2-\frac{3}{4}\ln3\bigg)
    + \frac{3}{4}\lmut^2\Bigg]\\&
     \eqindent{1}+\api^2\bigg(
    5.455\,\ccf^2 + 0.1028\,\cca\ccf
    -\big(1.078\,\nl+6.411 \,\nh\big)\ccf\ctr
    \\&\eqindent{2}
    + \lmut\big[3.095\,\ccf^2 + 0.3964\,\cca\ccf
      - \big(0.1512\,\nl+3.151\,\nh\big)\ccf\ctr \big]\\&
    \eqindent{2}
    + \lmut^2\big[1.862\,\ccf^2 + 0.6510\,\cca\ccf - 0.07763 \,\ccf\ctr\nf \big]
    \\
    &\eqindent{2}+ \lmut^3\big[0.6563\,\ccf^2 + 0.1146\,\cca\ccf
      - 0.04167 \,\ccf\ctr\nf \big] \bigg) \Bigg\} + \mathcal{O}(\api^3)\,,
    \stepcounter{equation}\tag{\theequation}\label{eq::zetaS3}\\
\zeta_\text{S}(t)
  &=1  + \api\,\ccf\Big( - \frac{1}{2} - \ln2
    - \frac{3}{4}\ln3
    - \frac{3}{4}\lmut\Big)\\
  & \eqindent{1}+ \api^2\Bigg\{\frac{1}{16}c_\chi^{(2)}
  + \ccf^2\Big(\frac{1}{2} + \frac{1}{2}\zeta_2
  - \frac{3}{2}\ln2 + \frac{1}{4}\ln^22 + \frac{21}{8}\ln3
  + 3\text{Li}_2\left(1/4\right)\Big)\\
  & \eqindent{2}
  + \cca\ccf\Big( - \frac{197}{48} - \frac{7}{16}\zeta_2 - \frac{9}{2}\ln2
      + \frac{1}{4}\ln^22 + \frac{9}{2}\ln3
    + \frac{3}{2}\text{Li}_2\left(1/4\right)\Big)\\
    & \eqindent{2}
    + \ccf\ctr\nf\Big(\frac{5}{6} + \frac{1}{4}\zeta_2\Big)
    + \lmut\bigg[\ccf^2\Big(\frac{3}{4}\ln2 + \frac{9}{16}\ln3
    + \frac{9}{32}\Big)\\
    & \eqindent{3}
    + \cca\ccf\Big( - \frac{11}{12}\ln2 - \frac{11}{16}\ln3
    - \frac{47}{32}\Big) + \ccf\ctr\nf\Big(\frac{1}{3}\ln2 + \frac{1}{4}\ln3
    + \frac{3}{8}\Big)\bigg]\\
& \eqindent{2}+ \lmut^2\bigg(\frac{9}{32}\ccf^2 - \frac{11}{32}\cca\ccf +
    \frac{1}{8}\ccf\ctr\nf\bigg)\Bigg\} + \mathcal{O}(\api^3)\,,
    \stepcounter{equation}\tag{\theequation}\label{eq::zetaS}\\
    \zeta_\text{V}(t) &=1 + \api\,\ccf\Big(\frac{1}{8} - \ln2
    - \frac{3}{4}\ln3\Big)\\
    & \eqindent{1}
    + \api^2\Bigg\{\frac{1}{16}c_\chi^{(2)} + \ccf^2\Big( - \frac{41}{128}
    - \frac{5}{32}\zeta_2 + \frac{3}{8}\ln2 + \frac{1}{4}\ln^22
    - \frac{3}{32}\ln3 + \frac{3}{2}\text{Li}_2\left(1/4\right)\Big)\\
    & \eqindent{2}
    + \cca\ccf\Big( - \frac{763}{384} - \frac{5}{32}\zeta_2 - \frac{13}{4}\ln2
    + \frac{1}{4}\ln^22 + \frac{27}{8}\ln3
    + \frac{21}{16}\text{Li}_2\left(1/4\right)\Big)\\
    & \eqindent{2}
    + \ccf\ctr\nf\Big(\frac{35}{96} + \frac{1}{8}\zeta_2\Big)
    + \lmut\bigg[\cca\ccf\Big( - \frac{11}{12}\ln2
      - \frac{11}{16}\ln3 + \frac{11}{96}\Big)
      \\&\eqindent{3}
    + \ccf\ctr\nf\Big(\frac{1}{3}\ln2 + \frac{1}{4}\ln3
    - \frac{1}{24}\Big)\bigg]\Bigg\} + \mathcal{O}(\api^3)\,,
    \stepcounter{equation}\tag{\theequation}\label{eq::zetaV}\\
    \zeta_\text{T}(t) &=1 + \api\,\ccf\Big( - \ln2 - \frac{3}{4}\ln3
    + \frac{1}{4}\lmut\Big)\\
    & \eqindent{1}
    + \api^2\Bigg\{\frac{1}{16}c_\chi^{(2)} + \ccf^2\Big( - \frac{7}{12}
- \frac{1}{4}\zeta_2 + \frac{4}{3}\ln2 + \frac{1}{4}\ln^22 - \frac{3}{4}\ln3
+ \text{Li}_2\left(1/4\right)\Big)\\
& \eqindent{2}
+ \cca\ccf\Big( - \frac{1159}{864} - \frac{1}{16}\zeta_2 - \frac{17}{6}\ln2
+ \frac{1}{4}\ln^22 + 3\ln3 + \frac{5}{4}\text{Li}_2\left(1/4\right)\Big)\\
& \eqindent{2}+ \ccf\ctr\nf\Big(\frac{47}{216} + \frac{1}{12}\zeta_2\Big)
 + \lmut\bigg[\ccf^2\Big( - \frac{1}{4}\ln2 - \frac{3}{16}\ln3
- \frac{19}{32}\Big)\\
& \eqindent{3}+ \cca\ccf\Big( - \frac{11}{12}\ln2 - \frac{11}{16}\ln3
+ \frac{257}{288}\Big) + \ccf\ctr\nf\Big(\frac{1}{3}\ln2 + \frac{1}{4}\ln3
- \frac{13}{72}\Big)\bigg]\\
& \eqindent{2}+ \lmut^2\bigg(\frac{1}{32}\ccf^2 + \frac{11}{96}\cca\ccf
- \frac{1}{24}\ccf\ctr\nf\bigg)\Bigg\} + \mathcal{O}(\api^3)\,.
\stepcounter{equation}\tag{\theequation}\label{eq::zetaT}
\end{align*}
$c_\chi^{(2)}$ is associated with the ringed scheme and would not appear if
the fermions were renormalized in the $\msbar$ scheme. Its explicit form is
given in \cref{eq::kane,eq::cane}.  The results for $\zeta_\text{S}^{(1)}(t)$
and $\zeta_\text{S}(t)$ are already known
from refs. \cite{Luscher:2013cpa,Artz:2019bpr,
  Mereghetti:2021nkt,Borgulat:2022cqe,Hieda:2016lly}.\footnote{In \citere{Artz:2019bpr},
$\zeta_\text{S}^{(1)}(t)$ was called the quark condensate $\langle
\mathring{S}(t)\rangle$, while in \citere{Borgulat:2022cqe},
$\zeta_\mathrm{S}(t)$ was called $s_S$ and quoted only numerically for $\qcd$
with $\nf=1$. $\zeta_\mathrm{S}(t)$ was also calculated in the context of
\citere{Harlander:2018zpi}, but not published.}  For the sake of brevity, we quote the three-loop results with only four significant digits here. In the
ancillary file accompanying this paper, we provide the results with higher
numerical accuracy, as well as analytic expressions for the logarithmic terms
and the coefficient of $\ccf\ctr\nf$ of $\zeta_\text{S}^{(1)}(t)$
(see \cref{app::anc}).
Moreover, while all results in this paper are in the ringed scheme,
the ancillary files contain the matching coefficients also in the
$\msbar$ scheme of the flowed fermions.

%- }}}
%- {{{ Pseudo-parity currents

\paragraph{Pseudo-parity currents.}

For the pseudo-parity currents, one has to introduce a non-minimal
renormalization in order to restore the associated Ward identities in regular
\qcd\ for the non-singlet cases, and the correct anomaly in the singlet
axialvector case, which are broken by adopting \cref{eq::iaso} combined with
minimal subtraction. Therefore, we define the renormalization constants of
\cref{eq::heat} in these cases as
\begin{equation}\label{eq::food}
  \begin{aligned}
    Z_\mathrm{X}(\api) &= Z^{\msbar}_\mathrm{X}(\api) Z_{5,\text{X}}(\api)\,,\qquad
    \text{X}\in\{\text{A},\text{P}\}\,,
  \end{aligned}
\end{equation}
where the \msbar\ part is given in \cref{eq::fold,eq::anom2}. The
$Z_{5,\text{X}}$ are finite renormalization constants given
in \cref{eq::xxxx}. For the non-singlet cases, taking them into account is
actually equivalent to working with a naively anti-commuting $\gamma_5$
combined with minimal subtraction, which means that
\begin{equation}\label{eq::aoki}
  \begin{aligned}
    \zeta_\text{P}(t)&=\zeta_\text{S}(t) \,,&&&
    \zeta^\text{ns}_\text{A}(t)&=\zeta_\text{V}(t)\,.
  \end{aligned}
\end{equation}
We explicitly verified these relations by using \cref{eq::iaso} and the
corresponding renormalization constants of \cref{eq::food}. This provides a
strong validity check of our calculational setup.

For the renormalized triangle contribution of \cref{eq::anax}, on the other
hand, we find
\begin{equation}\label{eq::mmrras}
  \begin{aligned}
    \zeta^\Delta_\text{A}(t) &=
\api^2\ccf\ctr\Big( - \frac{3}{8} + \frac{1}{8}\zeta_2 + \frac{9}{2}\ln2
- \frac{9}{4}\ln3 \\&
\eqindent{1}
+ \frac{15}{8}\text{Li}_2\left(1/4\right)
 + \frac{3}{4}\lmut\Big) + \mathcal{O}(\api^3)\,.
  \end{aligned}
\end{equation}

%- }}}
%- {{{ Additional checks

\paragraph{Additional checks.}
In order to further corroborate the correctness of our results, we performed
all calculations in general $R_\xi$ gauge and confirmed that the dependence on
the gauge parameter drops out in the final result. The only exception to this check
is $\zeta^{(3)}(t)$, for which the calculation in $R_\xi$ gauge exceeds our
computing resources. Through \nlo{}, the matching coefficients
$\zeta_\mathrm{X}(t)$ were already computed
in \citeres{Endo:2015iea,Hieda:2016lly}, and we find full agreement after
fixing the erroneous finite renormalization\footnote{The ``$(-4)$'' in the
$\order{g^2}$ coefficient of Eq.\,(2.7) in \citere{Hieda:2016lly} should read
``$(-8)$''. This also affects a number of the subsequent equations
in \citere{Hieda:2016lly}. We would like to thank H.~Suzuki for clarifying
communications on this issue.} for $\zeta_\text{P}(t)$ and
$\zeta_\text{S}(t)$. For $\zeta_\text{T}(t)$, only the bare \nlo\ result is
provided in \citere{Hieda:2016lly}, and it agrees with our bare result.

%- }}}
%- {{{ Numerics:

\paragraph{Numerical results for $\boldsymbol{\zeta}_\mathrm{\mathbf{V}}\boldsymbol{(}\boldsymbol{t}\boldsymbol{)}$.}

\begin{figure}
  \begin{center}
    \begin{tabular}{cc}
      \raisebox{0em}{%
        \mbox{%
          \includegraphics[%
            clip,width=.45\textwidth]%
                          {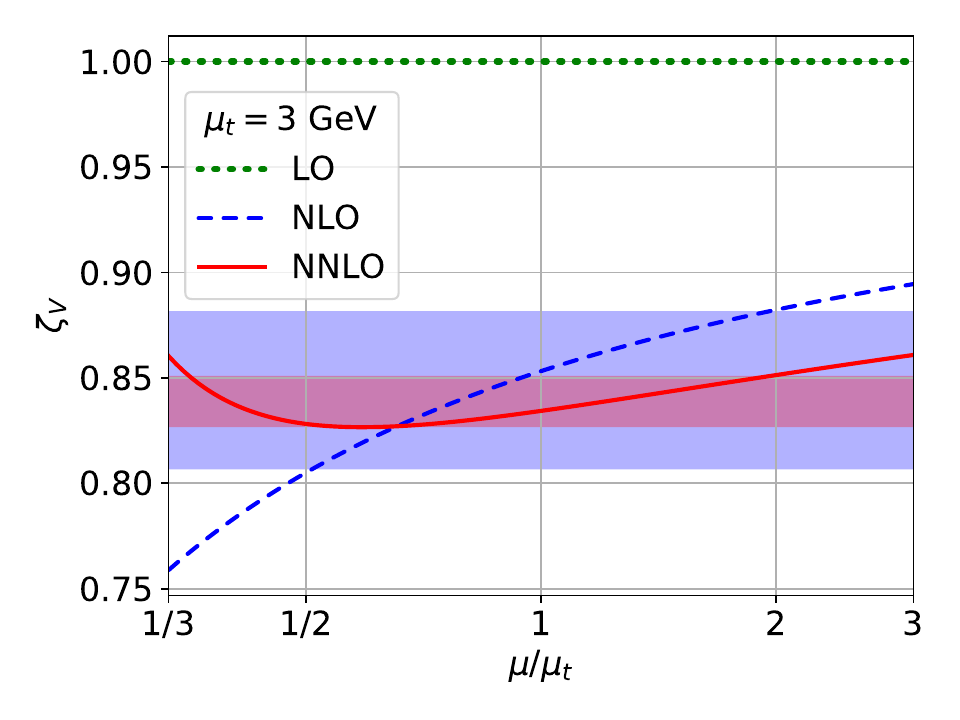}}}
      &
      \raisebox{0em}{%
        \mbox{%
          \includegraphics[%
            clip,width=.45\textwidth]%
                          {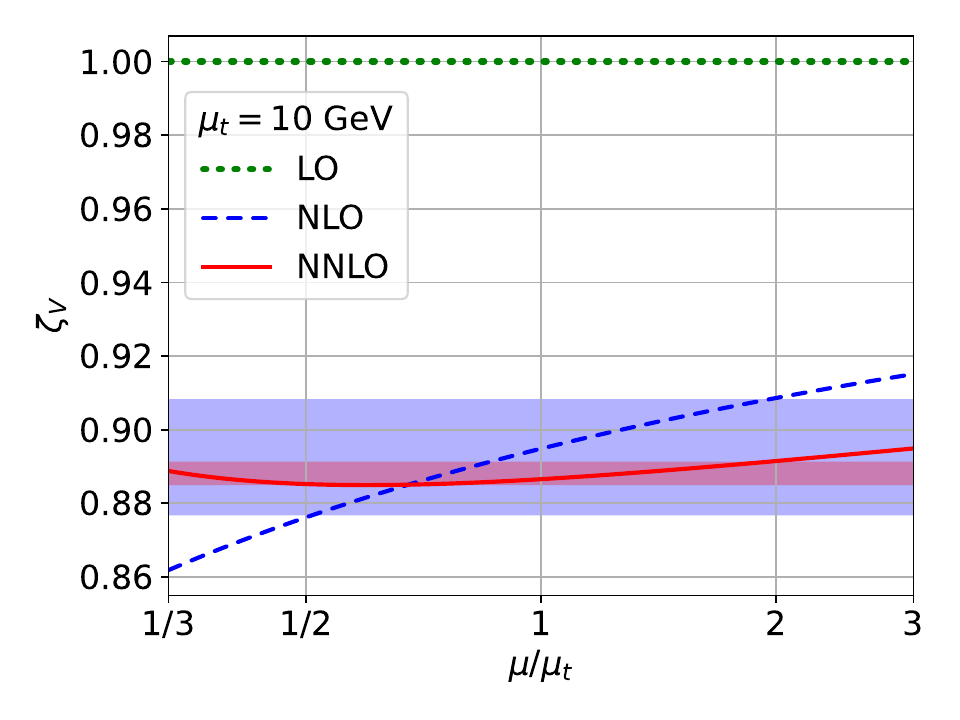}}}
    \end{tabular}
    \parbox{.9\textwidth}{
      \caption[]{\label{fig:results:zetaV}\sloppy The matching coefficient
        $\zeta_\text{V}(t)$ at two different values of
        $t=e^{-\EulerGamma}/(2\mu_t^2)$ as a function of $\mu/\mu_t$.  }}
  \end{center}
\end{figure}

In order to see the improvement of the impact of the \nnlo\ terms, we display
in \cref{fig:results:zetaV} the result for $\zeta_\text{V}(t)$ as a function of
the unphysical renormalization scale $\mu$ at \lo, \nlo, and \nnlo.  We
consider two exemplary values of the flow time $t$, defined in terms of the
energy scale $\mu_t$, see \cref{eq:calculation:fane}. The first one,
$\mu_t=3$\,GeV, corresponding to $t\approx 0.03$\,GeV$^{-2}$, seems
to constitute a reasonable compromise between the $t$-region preferred in
lattice calculations and the perturbative regime (see the preliminary study in \citere{Black:2023vju}).
For comparison, on the other hand, we also present results for
$\mu_t=10$\,GeV, corresponding to $t\approx 0.003$\,GeV$^{-2}$, which is well
in the perturbative regime.

For the plot at $\mu_t=3$\,GeV, we set $\nf=3$ and use
$\alpha_s^{(\nf=3)}(3\,\text{GeV}) = 0.2485$ as input.  For the plot at
$\mu_t=10$\,GeV, we set $\nf=5$ and use $\alpha_s^{(\nf=5)}(10\,\text{GeV}) =
0.1880$ as input.  We then evaluate $\alpha_s(\mu)$ through one-, two-, and
three-loop running for $\mu_t/3\leq \mu\leq 3\mu_t$ and insert it into
the \lo-, \nlo-, and \nnlo-approximation of $\zeta_\text{V}(t)$, respectively
(with the corresponding value for $\nf$), in order to obtain the three curves
in the plots.

%- {{{ fig:results:zetaV

%
%

%- }}}

Since this quantity is \abbrev{RG} invariant, we expect the $\mu$ dependence
to decrease from
\nlo\ to \nnlo, and this is indeed what we observe. Taking the variation of
$\mu$ around $\mu_t$ by a factor of two as an estimate of the perturbative
uncertainty, we find that it decreases from 4.4\% to 1.4\% at $\mu_t=3$\,GeV,
and from 1.8\% to 0.4\% at 10\,GeV. This is indicated by the red and blue
bands in the plot. Another important observation is that these bands overlap,
indicating that $\mu_t$ as defined in \cref{eq:calculation:fane} is indeed a
reasonable choice for the central renormalization scale.
Since the other matching coefficients are not \abbrev{RG} invariant, we refrain from showing the analogous plots for them.

%- }}}

%- }}}
%- {{{ section{Flowed anomalous dimension}\label{numerics}

\section{Flowed anomalous dimension}\label{sec:anom}

As suggested in \citere{Harlander:2020duo} for a general set of flowed
operators $\mathcal{O}=(\mathcal{O}_1,\ldots,\mathcal{O}_p)$, one may define
flowed anomalous dimensions which allow to resum their logarithmic
$t$-dependence. Let us briefly recapitulate the idea behind it. First consider
the flow time $t$ and the renormalization scale $\mu$ as independent
quantities. The regular operators $\mathcal{O}$ are then independent of $t$,
the flowed operators $\tilde{O}$ are independent of $\mu$, and the elements of
the matching matrix $\zeta$ are functions of $\api(\mu)$ and $\lmut$.
Therefore, neglecting terms that vanish as $t\to 0$,
\begin{equation}\label{eq:anom:airs}
  \begin{aligned}
    0 = t\dderiv{}{}{t}\mathcal{O} =
    t\dderiv{}{}{t}\zeta^{-1}\,\tilde{\mathcal{O}}
  \end{aligned}
\end{equation}
and thus
\begin{equation}\label{eq:anom:daut}
  \begin{aligned}
    t\dderiv{}{}{t} \tilde{\mathcal{O}} = \tilde\gamma\,
    \tilde{\mathcal{O}}\,,\quad\mbox{with}\quad \tilde\gamma =
    \left(t\dderiv{}{}{t}\zeta\right) \zeta^{-1}\,,
  \end{aligned}
\end{equation}
where the flowed anomalous dimension matrix $\tilde\gamma$ is a function of
$\api(\mu)$ and $\lmut$, which is, however, formally independent of $\mu$:
\begin{equation}\label{eq:anom:kick}
  \begin{aligned}
    \mu^2\dderiv{}{}{\mu^2}\tilde\gamma = 0\,.
  \end{aligned}
\end{equation}
Note that the derivative acting on $\zeta$ in \cref{eq:anom:daut} only affects
the logarithmic terms $\lmut$. The latter can be derived at higher orders by
noting that flowed operators (in the ringed scheme) are $\mu$-independent,
i.e.
\begin{equation}\label{eq:anom:heed}
  \begin{aligned}
    0 =
    \mu^2\dderiv{}{}{\mu^2}\tilde{\mathcal{O}} =
    \mu^2\dderiv{}{}{\mu^2}\zeta\, \mathcal{O} =
    \left[\left(\deriv{}{}{\lmut} + \api\beta\deriv{}{}{\api}\right)
    \zeta\right]\mathcal{O} + \zeta\,\gamma\,\mathcal{O}\,,
  \end{aligned}
\end{equation}
where $\gamma$ is the anomalous dimension matrix of the regular operators
$\mathcal{O}$, i.e.\ $\mu^2\dderiv{}{}{\mu^2}\mathcal{O} =
\gamma\mathcal{O}$, and $\beta$ is defined in \cref{sec:renormalization}.
Therefore,
\begin{equation}\label{eq:anom:dote}
  \begin{aligned} \tilde\gamma &=
    -\left(\api\beta\deriv{}{}{\api}\zeta\right)\zeta^{-1}
    - \zeta\gamma\zeta^{-1}\,.  \end{aligned}
\end{equation}
The term in brackets starts at $\order{\api^2}$, and thus the one-loop term of
the flowed anomalous dimension is given by the (negative of the) regular
anomalous dimension of the current. Furthermore, knowledge of $\zeta$ at order
$\api^n$ is sufficient to obtain $\tilde\gamma$ through order $\api^{n+1}$,
given that $\gamma$ is known to $\api^{n+1}$ and $\beta$ to $\api^{n}$.

It may be interesting to note that \cref{eq:anom:dote} can also be derived by
tying the flow time and the renormalization scale together from the start,
i.e., setting $\mu = c/\sqrt{t}$, with some constant $c$. In this case, also
the regular operators become $t$ dependent, while $\zeta$ depends on $t$ only
through $\api(c/\sqrt{t})$, and thus
\begin{equation}\label{eq:anom:heap}
  \begin{aligned}
    t\dderiv{}{}{t}\mathcal{O} = -\mu^2\dderiv{}{}{\mu^2}\mathcal{O}
    = -\gamma\mathcal{O}\quad\mbox{and}\quad
    t\dderiv{}{}{t}\zeta(t) = -\api\beta\deriv{}{}{\api}\zeta(t)\,,
  \end{aligned}
\end{equation}
which again leads to
\begin{equation}\label{eq:anom:eric}
  \begin{aligned}
    t\dderiv{}{}{t}\tilde{\mathcal{O}} =
    \left[-\api\beta\deriv{}{}{\api}\zeta-\zeta\gamma\right]\zeta^{-1}
    \tilde{\mathcal{O}}\,.
  \end{aligned}
\end{equation}

Applied to the current operators considered in this paper, \cref{eq:anom:daut}
reads
\begin{equation}\label{eq:anom:kane}
  \begin{aligned}
    t\dderiv{}{}{t}\tilde{j} =
    \tilde{\gamma}\,\tilde{j}\,,
  \end{aligned}
\end{equation}
and using
\cref{eq:anom:dote,eq::zetaS,eq::zetaV,eq::zetaT,eq::anom2,eq::anom3}, we
find
\allowdisplaybreaks
\begin{align*}
   \tilde\gamma_\text{S}(t)
    &=-\frac{3}{4}\api\ccf + \api^2\Bigg[-\frac{3}{32}\ccf^2 +
      \cca\ccf\Big(-\frac{47}{32}-\frac{11}{12}\ln2-\frac{11}{16}\ln3\Big)\\
      &\eqindent{2} + \ccf\ctr\nf\Big(\frac{3}{8} + \frac{1}{3}\ln2 +
      \frac{1}{4}\ln3\Big)\Bigg]\\ &
    \eqindent{1}+
    \api^3\Bigg\{\Big(\frac{11}{96}\cca-\frac{1}{24}\ctr\nf\Big)c_\chi^{(2)}
    -\frac{129}{128}\ccf^3
    + \ccf^2\cca\Big(\frac{305}{256} +
    \frac{11}{12}\zeta_2-\frac{11}{3}\ln2
    \\&\eqindent{3}-\frac{11}{24}\ln^22
    -\frac{11}{8}\ln2\ln3
    + \frac{33}{8}\ln3-\frac{33}{64}\ln^23 +
    \frac{11}{2}\text{Li}_2\left(1/4\right)\Big)\\ &
    \eqindent{2}+
    \ccf\cca^2\Big(-\frac{65869}{6912}-\frac{77}{96}\zeta_2-\frac{215}{24}\ln2
    + \frac{11}{24}\ln^22 + \frac{247}{32}\ln3 +
    \frac{11}{4}\text{Li}_2\left(1/4\right)\Big)\\ &
    \eqindent{2}+
    \ccf^2\ctr\nf\Big(\frac{19}{32}-\frac{1}{3}\zeta_2 + \frac{19}{12}\ln2 +
    \frac{1}{6}\ln^22 + \frac{1}{2}\ln2\ln3-\frac{21}{16}\ln3 +
    \frac{3}{16}\ln^23\\ &
    \eqindent{3}-2\text{Li}_2\left(1/4\right)-\frac{3}{4}\zeta_3\Big)
    + \ccf\ctr^2\nf^2\Big(-\frac{205}{432}-\frac{1}{6}\zeta_2\Big)\\ &
    \eqindent{2}+
    \ccf\cca\ctr\nf\Big(\frac{2071}{432} + \frac{3}{4}\zeta_2 +
    \frac{41}{12}\ln2-\frac{1}{6}\ln^22-\frac{43}{16}\ln3
    \\&\eqindent{3}-\text{Li}_2\left(1/4\right)
    + \frac{3}{4}\zeta_3\Big) \Bigg\} +
    \mathcal{O}(\api^4)\,,
  \stepcounter{equation}\tag{\theequation}\label{eq::gammatildeS}\\
  \tilde\gamma_\text{V}(t)
    &=\api^2\Bigg[\cca\ccf\Big(\frac{11}{96}-\frac{11}{12}\ln2-\frac{11}{16}\ln3\Big)
      + \ccf\ctr\nf\Big(-\frac{1}{24} + \frac{1}{3}\ln2 +
      \frac{1}{4}\ln3\Big)\Bigg]\\ &
    \eqindent{1}+
    \api^3\Bigg\{\Big(\frac{11}{96}\cca-\frac{1}{24}\ctr\nf\Big)c_\chi^{(2)} +
    \ccf^2\cca\Big(-\frac{77}{128}-\frac{55}{192}\zeta_2 +
    \frac{11}{12}\ln2\\ &
    \eqindent{3}
    -\frac{11}{24}\ln^22-\frac{11}{8}\ln2\ln3-\frac{33}{64}\ln^23
    + \frac{11}{4}\text{Li}_2\left(1/4\right)\Big)\\ &
    \eqindent{2}+
    \ccf\cca^2\Big(-\frac{8189}{2304}-\frac{55}{192}\zeta_2-\frac{20}{3}\ln2 +
    \frac{11}{24}\ln^22 + \frac{181}{32}\ln3 +
    \frac{77}{32}\text{Li}_2\left(1/4\right)\Big)\\ &
    \eqindent{2}+
    \ccf^2\ctr\nf\Big(\frac{3}{16} + \frac{5}{48}\zeta_2-\frac{1}{12}\ln2 +
    \frac{1}{6}\ln^22 + \frac{1}{2}\ln2\ln3 + \frac{3}{16}\ln3
    \\&\eqindent{3}+
    \frac{3}{16}\ln^23-\text{Li}_2\left(1/4\right)\Big)
    +\ccf\ctr^2\nf^2\Big(-\frac{35}{144}-\frac{1}{12}\zeta_2\Big)\\ &
    \eqindent{2}+
    \ccf\cca\ctr\nf\Big(\frac{559}{288} + \frac{1}{3}\zeta_2 +
    \frac{31}{12}\ln2-\frac{1}{6}\ln^22-\frac{31}{16}\ln3-\frac{7}{8}\text{Li}_2\left(1/4\right)\Big)\Bigg\}
    \\&\eqindent{1}+
    \mathcal{O}(\api^4)\,,
    \stepcounter{equation}\tag{\theequation}\label{eq::gammatildeV}\\
    \tilde\gamma_\text{T}(t)
    &=\frac{1}{4}\api\ccf + \api^2\bigg[-\frac{19}{32}\ccf^2 +
      \cca\ccf\Big(\frac{257}{288}-\frac{11}{12}\ln2-\frac{11}{16}\ln3\Big)\\ &
      \eqindent{2}+ \ccf\ctr\nf\Big(-\frac{13}{72} + \frac{1}{3}\ln2 +
      \frac{1}{4}\ln3\Big)\bigg]\\ &
    \eqindent{1}+
    \api^3\Bigg\{\Big(\frac{11}{96}\cca-\frac{1}{24}\ctr\nf\Big)c_\chi^{(2)} +
    \ccf^3\Big(\frac{365}{384}-\zeta_3\Big)\\ &
    \eqindent{2}+
    \ccf^2\cca\Big(-\frac{9287}{2304}-\frac{11}{24}\zeta_2 +
    \frac{22}{9}\ln2-\frac{11}{24}\ln^22\\ &
    \eqindent{3}-\frac{11}{8}\ln2\ln3-\frac{11}{8}\ln3-\frac{33}{64}\ln^23
    + \frac{11}{6}\text{Li}_2\left(1/4\right) + \frac{7}{4}\zeta_3\Big)\\ &
    \eqindent{1}
    +
    \ccf\cca^2\Big(-\frac{10079}{20736}-\frac{11}{96}\zeta_2-\frac{425}{72}\ln2
    + \frac{11}{24}\ln^22 + \frac{159}{32}\ln3
    \\&\eqindent{3}
    +\frac{55}{24}\text{Li}_2\left(1/4\right)-\frac{5}{8}\zeta_3\Big)\\ &
    \eqindent{2}+
    \ccf^2\ctr\nf\Big(\frac{161}{288} + \frac{1}{6}\zeta_2-\frac{23}{36}\ln2 +
    \frac{1}{6}\ln^22 + \frac{1}{2}\ln2\ln3 + \frac{11}{16}\ln3 +
    \frac{3}{16}\ln^23\\ &
    \eqindent{3}-\frac{2}{3}\text{Li}_2\left(1/4\right) +
    \frac{1}{4}\zeta_3\Big) +
    \ccf\ctr^2\nf^2\Big(-\frac{215}{1296}-\frac{1}{18}\zeta_2\Big)\\ &
    \eqindent{2}+\ccf\cca\ctr\nf\Big(\frac{923}{1296} + \frac{7}{36}\zeta_2 +
    \frac{83}{36}\ln2-\frac{1}{6}\ln^22-\frac{27}{16}\ln3
    \\&\eqindent{3}
    -\frac{5}{6}\text{Li}_2\left(1/4\right)-\frac{1}{4}\zeta_3\Big)
    \Bigg\} +
    \mathcal{O}(\api^4)\,,
    \stepcounter{equation}\tag{\theequation}\label{eq::gammatildeT}\\
 \tilde\gamma_\text{P}(t)
    &=
    \tilde\gamma_\text{S}(t)\,,
    \stepcounter{equation}\tag{\theequation}\label{eq::gammatildeP}\\
    \tilde\gamma_\text{A}^\text{ns}(t)
    &=
    \tilde\gamma_\text{V}(t)\,,
    \stepcounter{equation}\tag{\theequation}\label{eq::gammatildeAns}\\
    \tilde\gamma^\text{s}_\text{A}(t)
    &=\tilde{\gamma}^\text{ns}_\text{A}(t)
    + \frac{3}{4}\api^2\ccf\ctr\nf + \api^3\bigg[-\frac{9}{16}\ccf^2\ctr\nf
      \\&\eqindent{2}+
      \ccf\ctr^2\nf^2\Big(\frac{1}{6}-\frac{1}{12}\zeta_2-3\ln2 +
      \frac{3}{2}\ln3-\frac{5}{4}\text{Li}_2\left(1/4\right)\Big)\\ &
      \eqindent{2}+
      \ccf\cca\ctr\nf\Big(\frac{19}{24} + \frac{11}{48}\zeta_2 +
      \frac{33}{4}\ln2-\frac{33}{8}\ln3 +
      \frac{55}{16}\text{Li}_2\left(1/4\right)\Big) \bigg]
    \\&\eqindent{1}
    + \mathcal{O}(\api^4)\,.
    \stepcounter{equation}\tag{\theequation}
    \label{eq::gammatildeAs}
\end{align*}
Note that, in these formulas, $\api$ is still renormalized in the \msbar\
scheme, and we have set $\mu=\mu_t$, see \cref{eq:calculation:fane} (the
expression for general $\mu$ can be easily reconstructed
using \cref{eq:anom:kick}; it is also given in the ancillary file accompanying
this paper, see \cref{app::anc}). In order to eliminate any reference to
the \msbar\ scheme, one can simply convert $\api$ in these expressions to the
gradient-flow scheme according to~\cite{Luscher:2010iy}
\begin{equation}\label{eq:anom:fish}
  \begin{aligned}
    \api = \apigf\left[ 1 - e_{1}\apigf +
      \apigf^2(2e_{1}^2-e_{2}) + \order{\apigf^3}
      \ldots\right]\,,
  \end{aligned}
\end{equation}
where
\begin{equation}
  \begin{split}
    e_0 &= e_{00}\,,\qquad
    e_1= e_{10} + \beta_0\,\lmut\,,\\
    e_2 &= e_{20} + (2\beta_0\,e_{10}+\beta_1)\,\lmut+
    \beta_0^2\,\lmut^2\,,
    \label{eq:efull}
  \end{split}
\end{equation}
with $\beta_0$, $\beta_1$ from \cref{eq::anom}, and
\cite{Luscher:2010iy,Harlander:2016vzb,Artz:2019bpr}
\begin{equation}
  \begin{split}
    e_{00} &= 1\,,\qquad
    e_{10} = \left(\frac{13}{9} + \frac{11}{6} \ln 2 - \frac{3}{4} \ln 3\right)
    \cca - \frac{2}{9} \ctr\nf\,,\\
    e_{20} &= 1.74865\, \cca^2  - (1.97283\ldots)\, \cca\ctr\nf
    + \left(\zeta_3 - \frac{43}{48}\right)
    \ccf\ctr\nf \\&\eqindent{1}+ \left(\frac{1}{9}\zeta_2  -
    \frac{5}{81}\right) \ctr^2\nf^2\,.
    \label{eq:econst}
  \end{split}
\end{equation}
The exact expression for the coefficient of $\cca\ctr\nf$ can be found in
\citere{Artz:2019bpr}.

\begin{figure}
  \begin{center}
    \begin{tabular}{cc}
      \raisebox{0em}{%
        \mbox{%
          \includegraphics[%
            clip,width=.45\textwidth]%
                          {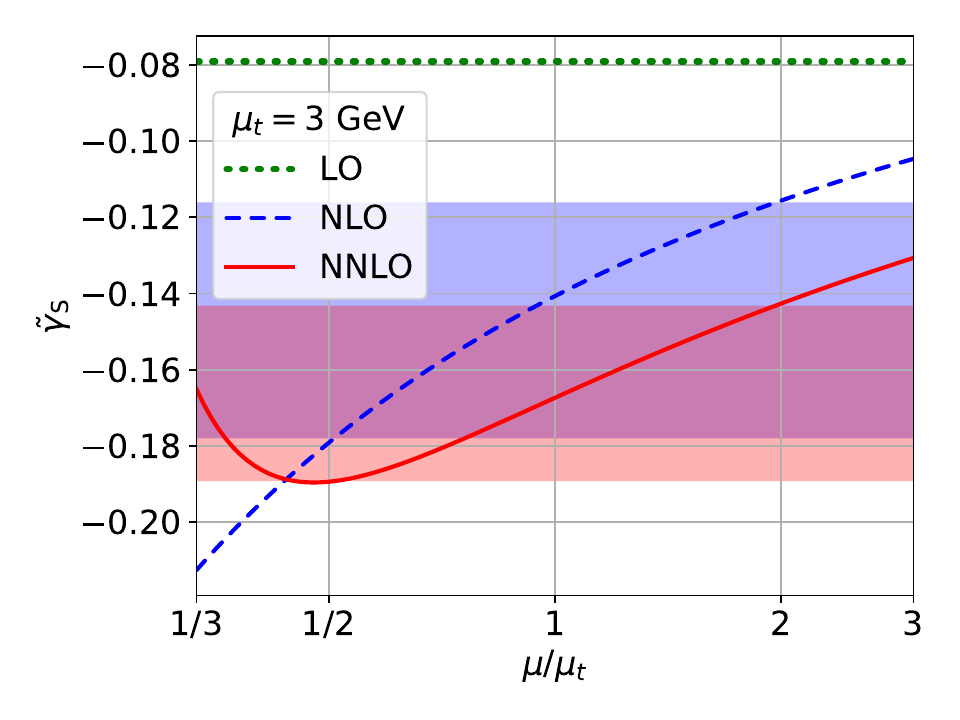}}}
      &
      \raisebox{0em}{%
        \mbox{%
          \includegraphics[%
            clip,width=.45\textwidth]%
                          {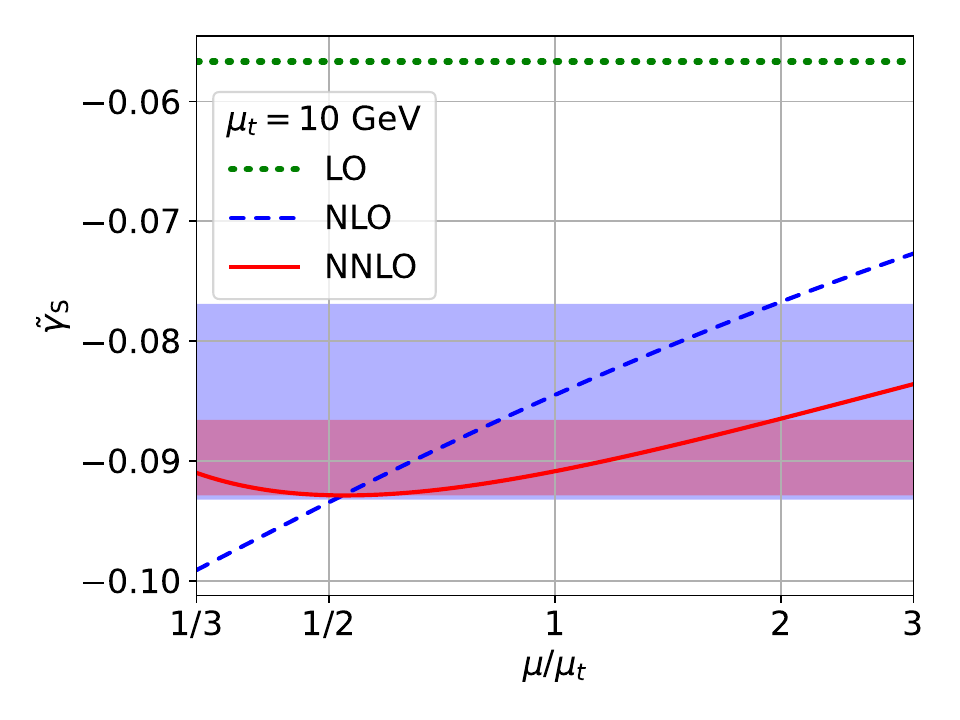}}}\\
      \raisebox{0em}{%
        \mbox{%
          \includegraphics[%
            clip,width=.45\textwidth]%
                          {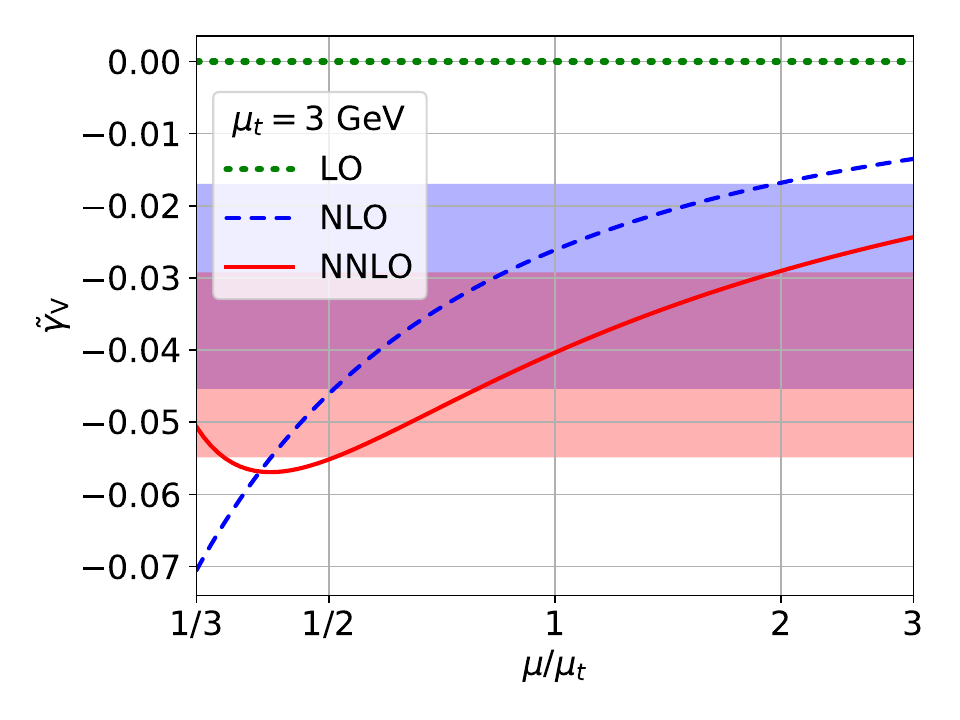}}}
      &
      \raisebox{0em}{%
        \mbox{%
          \includegraphics[%
            clip,width=.45\textwidth]%
                          {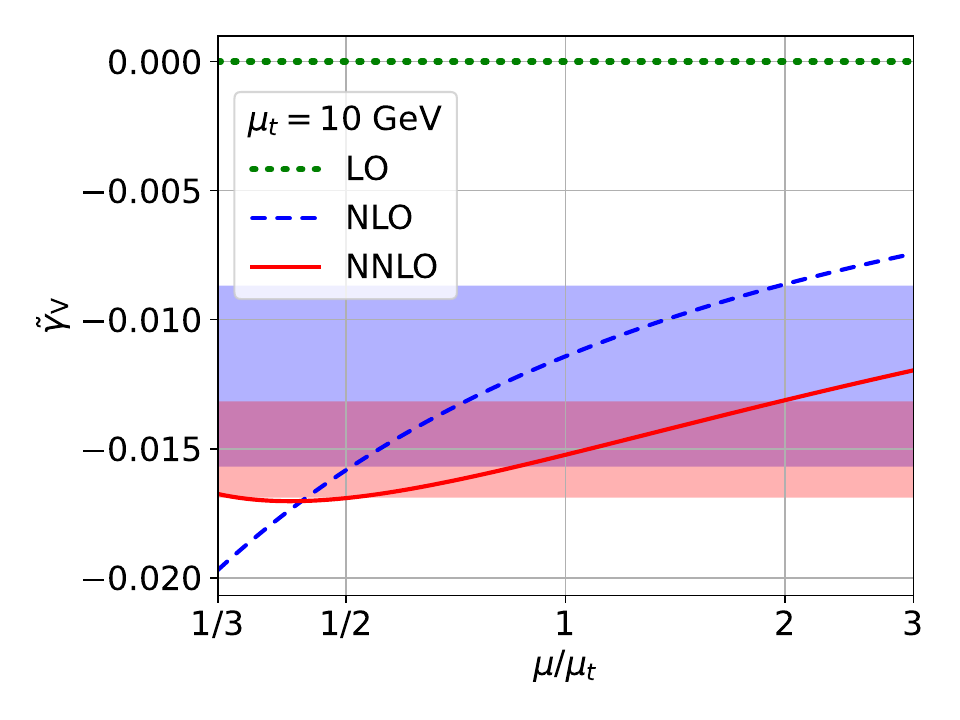}}}\\
      \raisebox{0em}{%
        \mbox{%
          \includegraphics[%
            clip,width=.45\textwidth]%
                          {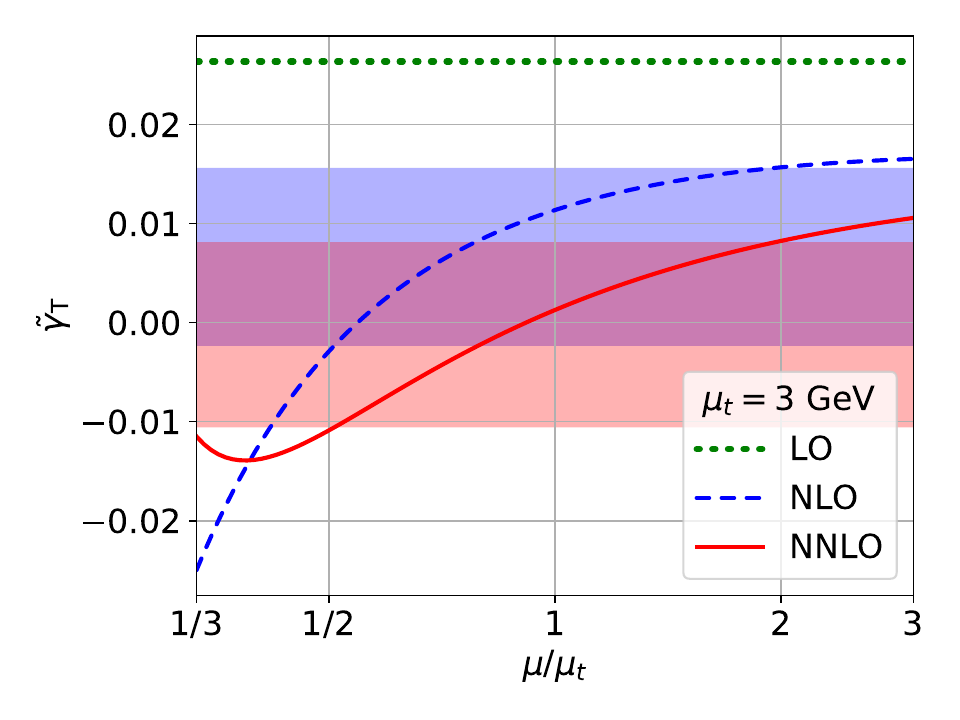}}}
      &
      \raisebox{0em}{%
        \mbox{%
          \includegraphics[%
            clip,width=.45\textwidth]%
                          {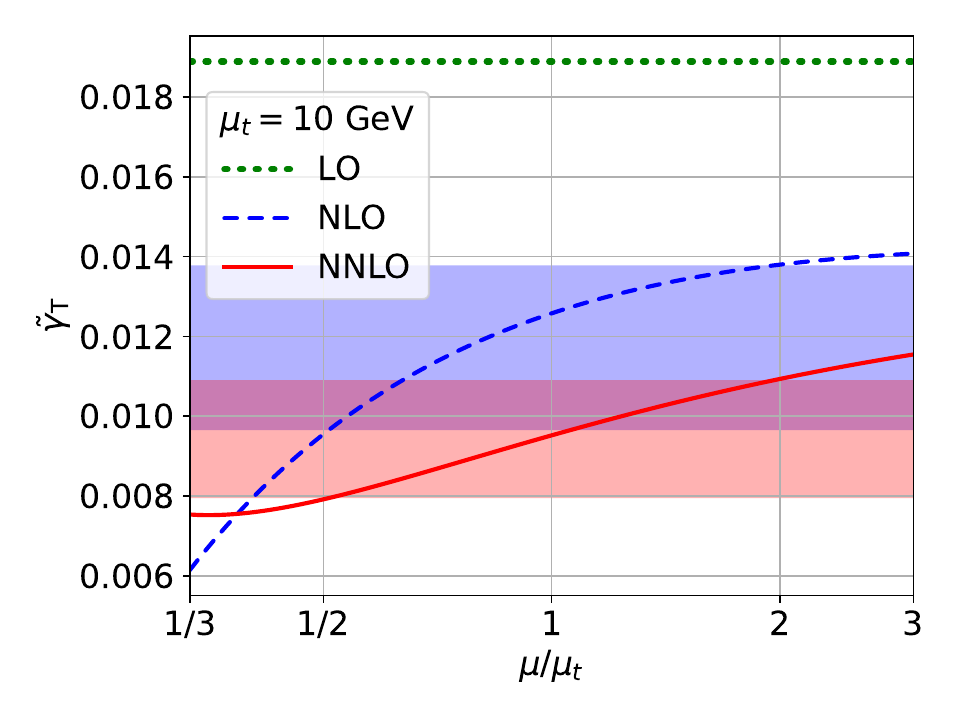}}}\\
      \raisebox{0em}{%
        \mbox{%
          \includegraphics[%
            clip,width=.45\textwidth]%
                          {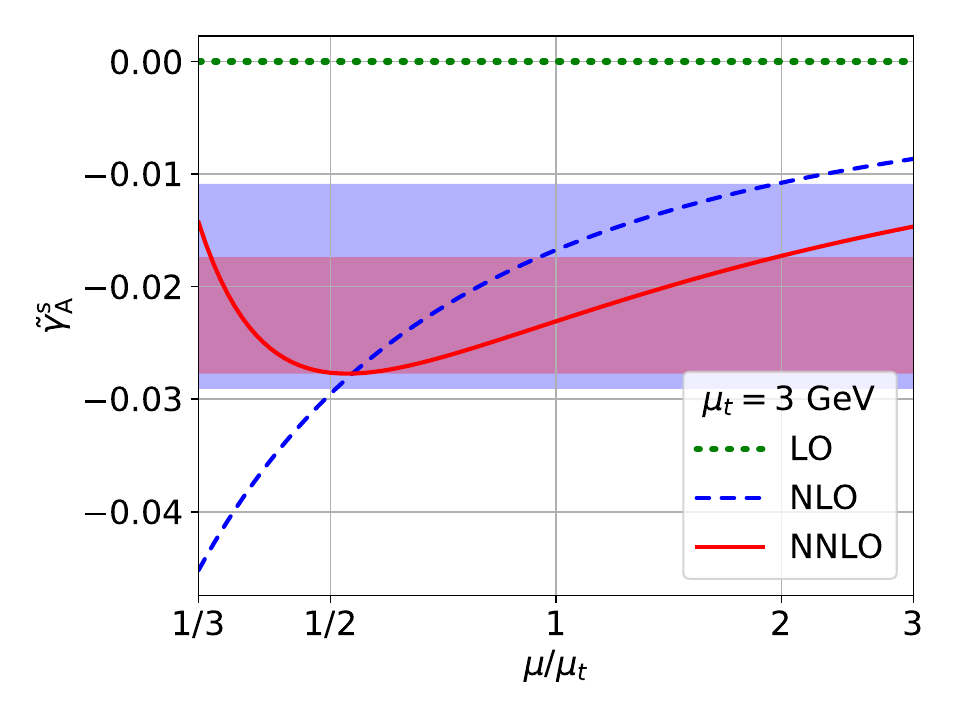}}}
      &
      \raisebox{0em}{%
        \mbox{%
          \includegraphics[%
            clip,width=.45\textwidth]%
                          {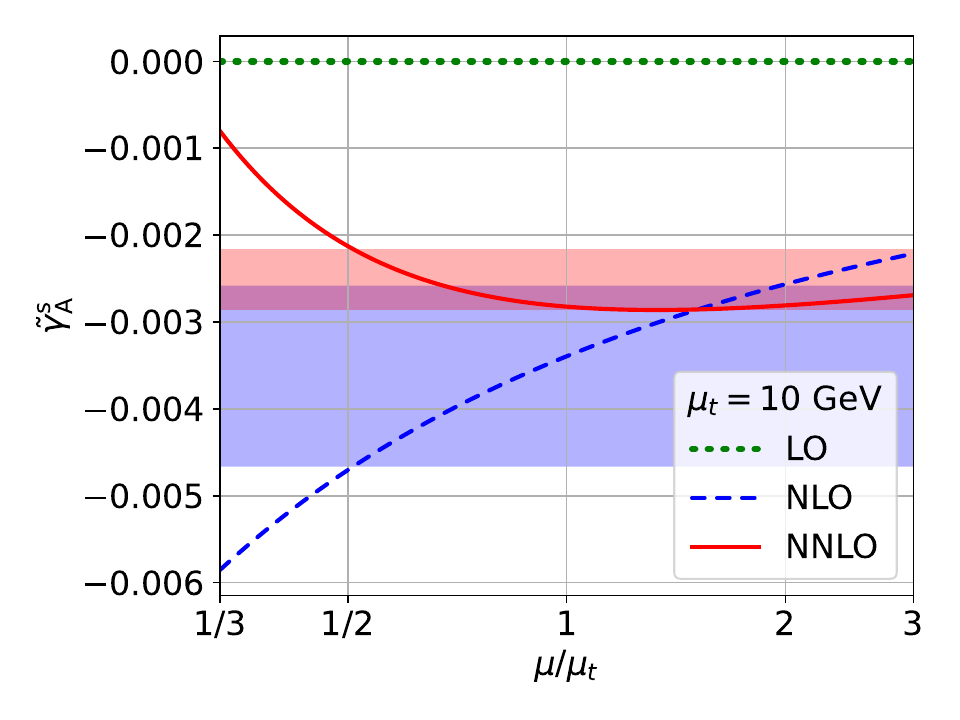}}}\\
    \end{tabular}
    \parbox{.9\textwidth}{
      \caption[]{\label{fig::gammaflow}\sloppy The flowed anomalous dimensions
        at two different values of $t=e^{-\EulerGamma}/(2\mu_t^2)$ as
        functions of $\mu/\mu_t$.  }}
  \end{center}
\end{figure}

Numerical values for the flowed anomalous dimensions are shown in
\cref{fig::gammaflow}. The input parameters are the same as in
\cref{fig:results:zetaV}. Also here, we observe the expected reduction of the
renormalization scale dependence when including higher orders, albeit
sometimes less pronounced than for $\zeta_\text{V}(t)$. But also here the
\nlo\ and the \nnlo\ uncertainty bands nicely overlap.

%- {{{ fig::gammaflow

%
%

%- }}}

%- }}}
%- {{{ Conclusions:

\section{Conclusions}
\label{sec:conclusions}

In this paper, we have considered the \sftx{} of the scalar, pseudoscalar,
vector, axialvector, and tensor currents and computed the corresponding
matching coefficients through \nnlo{} in \qcd{}. Possible applications of
these results are the calculation of the chiral condensate on the
lattice\,\cite{Taniguchi:2016ofw} or the semileptonic contributions to the
neutron electric dipole moment\,\cite{Buhler:2023gsg}.

Our results could also serve as alternatives to the ringed renormalization
scheme\,\cite{Makino:2014taa}, which requires the calculation of the vacuum
expectation value of the quark kinetic operator. In certain cases, it may be
more efficient to normalize quark matrix elements to one of the currents
instead. We believe that this strategy will especially find applications in
flavor physics, in particular in combination with the \sftx\ of the relevant
four-quark operators~\cite{Suzuki:2020zue,Harlander:2022tgk}.  A first
preliminary study, already employing the result for the axialvector current
obtained in the present paper, has been published in \citere{Black:2023vju}.

Finally, the simplicity of the quark currents could also be advantageous for
systematic studies of the \sftx{}, building up on the preliminary studies of
\citeres{Suzuki:2021tlr,Kim:2021qae,Hasenfratz:2022wll} in different contexts.

%- }}}
%- {{{ Acknowledgments:

\paragraph{Acknowledgments.}
We thank Matthew Black, Mattia Dalla Brida, Antonio Rago, Andrea Shindler, and
Oliver Witzel for motivation of this project, helpful discussions, and
comments on the manuscript.

This research was supported by the \textit{Deutsche Forschungsgemeinschaft
  (DFG, German Research Foundation) under grant 396021762 - TRR 257}.  The
work of F.L.~was supported by the Swiss National Science Foundation (SNSF)
under contract \href{https://data.snf.ch/grants/grant/211209}{TMSGI2\_211209}.

%- }}}
%- {{{ appendix:

\begin{appendix}

%- {{{ section{Renormalization constants}

\section{Renormalization constants}\label{sec:renormalization}
  In this appendix, we collect the renormalization constants required to
  arrive at the finite results presented in this paper.  Most of these
  constants are known to higher loop orders, but we will display only the
  orders which are relevant for our calculation.

  The relation between the bare and \msbar-renormalized gauge coupling is
  given by the regular-\qcd\ expression
\begin{equation}\label{eq::gren}
  \begin{aligned}
    g_\bare=\left(\frac{\mu^2e^{\EulerGamma}}{4\pi}\right)^{\epsilon/2}
    Z_g(\api(\mu))g(\mu)\,,
  \end{aligned}
\end{equation}
with $\mu$ the renormalization scale, $\EulerGamma = 0.577215\ldots$ the
Euler-Mascheroni constant,
\begin{equation}\label{eq::geno}
  \begin{aligned}
    Z_g(\api) &= 1 -
    \api\frac{\beta_0}{2\epsilon}  +
    \api^2\left(\frac{3\beta_0^2}{8\epsilon^2}
    -\frac{\beta_1}{4\epsilon}\right)  +
\mathcal{O}(\api^3)\,,
  \end{aligned}
\end{equation}
$\api$ from \cref{eq:results:acis}, and the coefficients of the \qcd\ beta
function,
\begin{equation}\label{eq:ren:kiss}
  \begin{aligned}
    \beta(\api) = -\ep-\api\sum_{n=0}^\infty\beta_n\api^n\,,
  \end{aligned}
\end{equation}
given by
\begin{equation}\label{eq::anom}
  \begin{aligned}
    \beta_0 &= \frac{1}{4}\left(\frac{11}{3}\cca
    -\frac{4}{3}\ctr\nf\right)\,,&&&
    \beta_1 &= \frac{1}{16}\left[\frac{34}{3}\cca^2-\left(4\ccf
       + \frac{20}{3}\cca\right)\ctr\nf\right]\,.\\
  \end{aligned}
\end{equation}
The \qcd\ color factors are defined in \cref{eq::joss}, and $\nf$ is the
number of quark flavors.  The remaining \msbar\ renormalization constants are
cast into the generic form
\begin{equation}\label{eq::fold}
  \begin{aligned}
 Z^{\msbar}(\api) &= 1 - \api\,\frac{\gamma_0}{\epsilon}  +
 \api^2\left[\frac{1}{2\epsilon^2}\left(\gamma^2_{0}  +
   \beta_0\gamma_{0}\right)
   -\frac{\gamma_{1}}{2\epsilon}\right]  +
 \mathcal{O}(\api^3)\,,
  \end{aligned}
\end{equation}
where the $\gamma_n$ are the perturbative coefficients of the corresponding
anomalous dimensions. For the quark mass, the latter is defined as
\begin{equation}\label{eq:ren:iowa}
  \begin{aligned}
    \gamma_m(\api) = -\api\sum_{n\geq 0}\api^n\gamma_{m,n}
    \equiv
    -\api\beta(\api)\dderiv{}{}{\api} \ln Z_m^{\msbar}(\api)\,,
  \end{aligned}
\end{equation}
and thus
\begin{equation}\label{eq:ren:jiva}
  \begin{aligned}
    \mu^2\dderiv{}{}{\mu^2}m(\mu) = \gamma_m(\api)m(\mu)\,.
  \end{aligned}
\end{equation}
For the current $j$, we define
\begin{equation}\label{eq:ren:kiev}
  \begin{aligned}
    \gamma(\api) = \api\sum_{n\geq 0}\api^n\gamma_{n} \equiv
    \api\beta(\api)\dderiv{}{}{\api} \ln Z^{\msbar}(\api)\,.
  \end{aligned}
\end{equation}
The renormalization group equation for the currents is thus given by
\begin{equation}\label{eq:ren:haem}
  \begin{aligned}
    \mu^2\dderiv{}{}{\mu^2}j(\mu) =
    \left[\gamma(\api)+\gamma^\text{fin}(\api)\right]
    j(\mu)\,,
  \end{aligned}
\end{equation}
where $\gamma^\text{fin}$ arises from any finite renormalization as
introduced for the pseudo-parity currents when adopting
\cref{eq::iaso}. Specifically, if
\begin{equation}\label{eq:ren:cara}
  \begin{aligned}
Z(\api)=Z^\text{fin}(\api)\,Z^{\msbar}(\api)\,,\quad\text{with}\quad
Z^\text{fin}(\api) &= 1+\sum_{n=1}^\infty \api^n z_{n0}\,,
  \end{aligned}
\end{equation}
then
\begin{equation}\label{eq:ren:able}
  \begin{aligned}
    \gamma^\text{fin}(\api) &= -\api^2\,\beta_0\,z_{10}
    - \api^3\bigg[\beta_1\,z_{10} - \beta_0\,z_{10}^2
      + 2\beta_0\,z_{20}\bigg] + \order{\api^4}\,.
  \end{aligned}
\end{equation}

In this paper, we need the \msbar\ quark mass renormalization constant
$Z_m\equiv Z_m^{\msbar}$
through $\order{\api^3}$, given by
%\begin{equation}\label{eq:ren:jeer}
%  \begin{aligned}
%    \gamma_{m,0} &= \frac{3}{4}\ccf\,,&&&
%    \gamma_{m,1} &= \frac{3}{32}\ccf^2+\frac{97}{96}\cca\ccf
%   -\frac{5}{24}\ccf\ctr\nf\,,
%  \end{aligned}
%\end{equation}
\begin{equation}\label{eq:ren:jeer}
  \begin{aligned}
    \gamma_{m,0} &= \frac{3}{4}\ccf\,,\qquad
    \gamma_{m,1} = \frac{3}{32}\ccf^2
    + \frac{97}{96}\cca\ccf
    - \frac{5}{24}\ccf\ctr\nf\,,\\
    \gamma_{m,2} &= \frac{1}{64}\bigg[
      \frac{129}{2}\ccf^3
      - \frac{129}{4}\ccf^2\cca
      +\frac{11413}{108}\ccf\cca^2
      \\&
      + \ccf^2\ctr\nf\left(
      - 46
      + 48\zeta_3
      \right)
      + \ccf\cca\ctr\nf\left(
      - \frac{556}{27}
      - 48\zeta_3
      \right)
      -\frac{140}{27}\ccf\ctr^2\nf^2
      \bigg]\,,
  \end{aligned}
\end{equation}
as well as the renormalization constant of the vacuum energy $Z_0$ through $\mathcal{O}(\api^2)$. It is related to the corresponding anomalous dimension $\gamma_0$ through
\begin{equation}
\gamma_0(\api) = \left[4\gamma_m(\api)-\epsilon\right]Z_0(\api)+\beta(\api)\api\frac{\partial}{\partial\api}Z_0(\api) \equiv -\frac{\nc\nh}{(4\pi)^2}\sum_{n\geq 0}\api^n\gamma_{0,n} \, ,
\end{equation}
which leads to
\begin{equation}
  \begin{aligned}\label{eq::z0}
  Z_0(\api) = \frac{\nc\nh}{(4\pi)^2\epsilon}
  \Bigg\{ 1 + \api &\left(\frac{\gamma_{0,1}}{2}-\frac{2\gamma_{m,0}}{\epsilon}\right)
  + \api^2\Bigg[ \frac{2}{3\epsilon^2}(\beta_0\gamma_{m,0}+4\gamma_{m,0}^2) \\
  &-\frac{1}{6\epsilon}(\beta_0\gamma_{0,1}
  +4\gamma_{0,1}\gamma_{m,0}+8\gamma_{m,1})
  +\frac{1}{3}\gamma_{0,2}\Bigg]\Bigg\} + \mathcal{O}(\api^3) \, .
  \end{aligned}
\end{equation}
The first three perturbative coefficients are given by
\cite{Spiridonov:1988md,Chetyrkin:1994ex}
\begin{equation}
\begin{aligned}
  &\gamma_{0,0}=1 \,,\qquad \gamma_{0,1}=\ccf \,,\\
  &\gamma_{0,2}=-\ccf^2\left(\frac{131}{32}-3\zeta_3\right)-\ccf\cca\left(-\frac{109}{32}+\frac{3}{2}\zeta_3\right)-\ccf\ctr\left(\frac{5}{8}\nf+3\nh\right) \,.
\end{aligned}
\end{equation}

The current renormalizations
$Z^{\msbar}_\mathrm{T}$,
$Z^{\msbar}_\mathrm{P}$, and
$Z^{\msbar}_\mathrm{A}$
are needed through $\order{\api^2}$. For these\,\cite{Broadhurst:1994se,Larin:1993tq,Larin:1991tj},
\begin{equation}\label{eq::anom2}
  \begin{aligned}
    \gamma_{\text{T},0} &=
    - \frac{1}{4}\ccf\,,\qquad &
    \gamma_{\text{T},1} &=
    \frac{19}{32}\ccf^2
    - \frac{257}{288}\cca\ccf
    + \frac{13}{72}\ccf\ctr\nf\,,\\
    \gamma_{\text{P},0} &= \frac{3}{4}\ccf\,,\qquad &
    \gamma_{\text{P},1} &=
    \frac{3}{32}\ccf^2
    - \frac{79}{96}\cca\ccf
    + \frac{11}{24}\ccf\ctr\nf\,,\\
    \gamma^\text{ns}_{\text{A},0} &=
    0\,,\qquad &
    \gamma^\text{ns}_{\text{A},1} &=
    - \frac{11}{12}\cca\ccf
    + \frac{1}{3}\ccf\ctr\nf\,,\\
    \gamma^\text{s}_{\text{A},0}
    &= 0\,,\qquad &
    \gamma^\text{s}_{\text{A},1} &=
    \gamma^\text{ns}_{\text{A},1}
    - \frac{3}{4}\ccf\ctr\nf\,.
  \end{aligned}
\end{equation}
In order to derive the $\mathcal{O}(\api^3)$ terms of $\tilde{\gamma}$ in
\cref{sec:anom}, we also need the terms at $\order{\api^3}$\,\cite{Gracey:2000am,Larin:1993tq,Larin:1991tj}:
\begin{equation}\label{eq::anom3}
  \begin{aligned}
    \gamma_{\text{T},2} &= \left(
    \zeta_3
    -\frac{365}{384}
    \right)\ccf^3
    + \left(
    \frac{6823}{2304}
    - \frac{7}{4}\zeta_3
    \right)\ccf^2\cca
    + \left(
    \frac{5}{8}\zeta_3
    - \frac{13639}{6912}
    \right)\ccf\cca^2\\
       &\eqindent{1}-\left(
    \frac{49}{288}
    + \frac{1}{4}\zeta_3
    \right)\ccf^2\ctr\nf
    + \frac{1}{48}\ccf\ctr^2\nf^2
    + \left(
    \frac{251}{432}
    +\frac{1}{4}\zeta_3
    \right)\ccf\cca\ctr\nf\,,\\
    \gamma_{\text{P},2} &=
\frac{599}{2304}\cca^2\ccf
- \frac{3203}{768}\cca\ccf^2
+ \frac{129}{128}\ccf^3 +
 \frac{29}{48}\cca\ccf\nf\ctr
 + \frac{107}{96}\ccf^2\nf\ctr
 \\&\eqindent{1}
 - \frac{17}{144}\ccf\ctr^2\nf^2
 - \frac{3}{4}\zeta_3\cca\ccf\ctr\nf
 + \frac{3}{4}\zeta_3\ccf^2\ctr\nf\,,\\
    \gamma^\text{ns}_{\text{A},2} &=
    \frac{77}{48}\ccf^2\cca
    - \frac{1789}{864}\ccf\cca^2
    - \frac{1}{3}\ccf^2\ctr\nf
    - \frac{1}{54}\ccf\ctr^2\nf^2
    + \frac{26}{27}\ccf\cca\ctr\nf\,,\\
    \gamma^\text{s}_{\text{A},2} &=
    \gamma^\text{ns}_{\text{A},2}
    + \frac{9}{16}\ccf^2\ctr\nf
    - \frac{1}{24}\ccf\ctr^2\nf^2
    - \frac{109}{96}\ccf\cca\ctr\nf\,.
  \end{aligned}
\end{equation}
Recall that the vector current does not require renormalization, and the
scalar current renormalizes with $Z_\text{S}\equiv Z_m$. The finite
renormalization constants for the axial and the pseudoscalar current
introduced in \cref{eq::food} are given by~\cite{Larin:1993tq,Larin:1991tj}
\begin{equation}\label{eq::xxxx}
  \begin{aligned}
    Z_{5,\text{P}}(\api) &= 1 -2\,\api\ccf +
    \api^2\Big( \frac{1}{72}\ccf\cca+\frac{1}{18}\ccf\ctr\nf\Big) +
    \mathcal{O}(\api^3)\,,\\
    Z^\text{ns}_{5,\text{A}}(\api) &= 1 - \api\ccf
    + \api^2\left(\frac{11}{8}\ccf^2 -
    \frac{107}{144}\ccf\cca+\frac{1}{36}\ccf\ctr\nf\right)
    +\mathcal{O}(\api^3)\,,\\
    Z^\text{s}_\text{5,A}(\api) &= Z^\text{ns}_\text{5,A}
    +\frac{3}{16}\api^2\,\ccf\ctr\nf + \order{\api^3}\,.
  \end{aligned}
\end{equation}
Let us remark that quoting the results for $Z^\text{ns}_{5,\text{A}}$ and
$Z_{5,\text{P}}$ is redundant, because they could be derived from
\cref{eq:ren:able} and
\begin{equation}\label{eq:renormalization:coos}
  \begin{aligned}
    \gamma_\text{S} &= \gamma_\text{P} + \gamma_\text{P}^\text{fin}\,,\qquad
    \gamma_\text{V} = \gamma_\text{A}^\text{ns}
    + \gamma_\text{A}^\text{ns,fin}\,.
  \end{aligned}
\end{equation}

The flowed-quark field renormalization constant introduced in \cref{eq::zchi-def} assumes the same form as \cref{eq::fold}
with the anomalous dimensions given by\,\cite{Luscher:2013cpa,Harlander:2018zpi}
\begin{equation}\label{eq::duit}
  \begin{aligned}
    \gamma_{\chi,0} &= -\frac{3}{4}\ccf\,,&&&
    \gamma_{\chi,1} &= \left(\frac{1}{2}\ln 2-\frac{223}{96}\right)\cca\ccf
    +\left(\frac{3}{32}+\frac{1}{2}\ln 2\right)\ccf^2 +\frac{11}{24}\ccf\ctr\nf\,.
  \end{aligned}
\end{equation}
Besides the \msbar{} scheme, the so-called ringed scheme is determined from the all-order condition\,\cite{Makino:2014taa}
\begin{equation}\label{eq::adze}
  \begin{aligned}
    \mathring{Z}_\chi\langle\bar\chi_p(t)
    \overleftrightarrow{\slashed{\mathcal{D}}}^\text{F}\chi_p(t)\rangle\bigg|_{m=0}\equiv
    -\frac{2\nc\nf}{(4\pi t)^2}\,.
  \end{aligned}
\end{equation}
It is related to the \msbar{} scheme by
\begin{equation}\label{eq::city}
  \begin{aligned}
    \mathring{Z}_\chi &= \zeta_\chi(t,\mu)Z^{\msbar}_\chi\,,
  \end{aligned}
\end{equation}
with the finite renormalization constant\,\cite{Makino:2014taa,Artz:2019bpr}
\begin{equation}\label{eq::ivis}
  \begin{aligned}
    \zeta_\chi(t,\mu) = 1&
    -\api\left(\gamma_{\chi,0}\lmut
    +\frac{3}{4}\ccf\ln 3+\ccf\ln 2\right)\\
    &+\api^2\Bigg\{\frac{\gamma_{\chi,0}}{2}\left(\gamma_{\chi,0}-\beta_0
    \right)\lmut^2
    +\Big[\gamma_{\chi,0}\left(\beta_0
      -\gamma_{\chi,0}\right)\ln3\\
      &+\frac{4}{3}\gamma_{\chi,0}\left(\beta_0
      -\gamma_{\chi,0}\right)\ln 2-\gamma_{\chi,1}\Big]\lmut
    +\frac{c_\chi^{(2)}}{16}\Bigg\}+\mathcal{O}(\api^3)\,,
  \end{aligned}
\end{equation}
where
\begin{equation}\label{eq::kane}
  \begin{aligned}
    c_\chi^{(2)} &= \cca\ccf\, c_{\chi,\text{A}}
    +\ccf^2\, c_{\chi,\text{F}}+\ccf\ctr\nf\, c_{\chi,\text{R}}\,.
  \end{aligned}
\end{equation}
The coefficients have been evaluated in \citere{Artz:2019bpr}:\footnote{The
factor $-1/18$ should read $1/18$ in Eq.\,(B.3) of \citere{Artz:2019bpr}
(Eq.\,(130) in the \texttt{arXiv} version).}
\begin{equation}\label{eq::cane}
  \begin{aligned}
    c_{\chi,\text{A}} &= -23.7947,\qquad c_{\chi,\text{F}}= 30.3914,\qquad\\
    c_{\chi,\text{R}} &=
    -\frac{131}{18}+\frac{46}{3}\zeta_2
    +\frac{944}{9}\ln 2+\frac{160}{3}\ln^2 2
    -\frac{172}{3}\ln 3+\frac{104}{3}\ln2 \ln 3\\
    &-\frac{178}{3}\ln^2 3+\frac{8}{3}\text{Li}_2(1/9)
    -\frac{400}{3}\text{Li}_2(1/3)+\frac{112}{3}\text{Li}_2(3/4)
    =-3.92255\ldots\,.
  \end{aligned}
\end{equation}
Only digits are quoted in \cref{eq::cane} which are not affected by the
numerical uncertainty.

%- }}}

\section{Ancillary file}\label{app::anc}

For the reader's convenience, we provide the main results of this paper as an
ancillary file in \texttt{Mathematica} format.  The results are encoded in the
expressions listed in \cref{tab::results}. The matching coefficients $\zeta$
are provided both in the ringed scheme of the fermions as well as in the
$\msbar$ scheme. One may switch between the two schemes by setting the
variable \verb$Xzetachi$ to 0 ($\msbar$ scheme) or 1 (ringed scheme). The
flowed anomalous dimensions $\tilde{\gamma}$ are provided only in the ringed
scheme.

The results depend on the variables listed in \cref{tab::variables}.  The
matching coefficients in the ringed scheme also contain the symbol \verb$C2$,
which corresponds to the coefficient $c_\chi^{(2)}$, defined in
\cref{eq::kane,eq::cane}. The latter relations are provided in the form of a
\texttt{Mathematica} replacement rule named \texttt{ReplaceC2}.

\begin{table}
  \begin{center}
    \caption{\label{tab::results} The expressions of the ancillary file that
      encode the main results of this paper.}
    \begin{tabular}{lll}
      expression & meaning & reference\\\hline
      \verb$zetaS1$ &  $\zeta_\text{S}^{(1)}$ &  \cref{eq::zetaS1}\\
      \verb$zetaS3$ &  $\zeta_\text{S}^{(3)}$ &  \cref{eq::zetaS3}\\
      \verb$zetaS$ &  $\zeta_\text{S}$ &  \cref{eq::zetaS}\\
      \verb$zetaV$ &  $\zeta_\text{V}$ &  \cref{eq::zetaV}\\
      \verb$zetaT$ &  $\zeta_\text{T}$ &  \cref{eq::zetaT}\\
      \verb$zetaAns$ &  $\zeta_\text{A}^\text{ns}$ &  \cref{eq::aoki}\\
      \verb$zetaP$ &  $\zeta_\text{P}$ &  \cref{eq::aoki}\\
      \verb$zetaAtriangle$ &  $\zeta_\text{A}^\Delta$ &
      \cref{eq::mmrras}\\
      \verb$tildegammaS$ &  $\tilde{\gamma}_\text{S}$ &
      \cref{eq::gammatildeS}\\
      \verb$tildegammaV$ &  $\tilde{\gamma}_\text{V}$ &
      \cref{eq::gammatildeV}\\
      \verb$tildegammaT$ &  $\tilde{\gamma}_\text{T}$ &
      \cref{eq::gammatildeT}\\
      \verb$tildegammaAns$ &  $\tilde{\gamma}_\text{A}^\text{ns}$ &
      \cref{eq::gammatildeAns}\\
      \verb$tildegammaAs$ &  $\tilde{\gamma}_\text{A}^\text{s}$ &
      \cref{eq::gammatildeAs}\\
      \verb$tildegammaP$ &  $\tilde{\gamma}_\text{P}$ &
      \cref{eq::gammatildeP}
    \end{tabular}
  \end{center}
\end{table}

\begin{table}
  \begin{center}
    \caption{\label{tab::variables}Notation for the variables in the
      ancillary file.}
    \begin{tabular}{lll}
      symbol & meaning & reference\\\hline
      \verb$nc$ &  $\nc$ &  \cref{eq::joss} \\
      \verb$tr$ &  $\ctr$ &   \cref{eq::joss}\\
      \verb$cf$ &  $\ccf$ &  \cref{eq::joss}\\
      \verb$ca$ &  $\cca$ &  \cref{eq::joss}\\
      \verb$Lmut$ &  $\lmut$ &  \cref{eq:calculation:fane}\\
      \verb$as$ &$\api$ &  \cref{eq:results:acis}\\
      \verb$nf$ &  $\nf$ &  \cref{sec:currents}\\
      \verb$nl$ &  $\nl$ &  \cref{sec:currents}\\
      \verb$nh$ &  $\nh$ &  \cref{sec:currents}\\
    \end{tabular}
  \end{center}
\end{table}

\end{appendix}

%- }}}

%- }}} body:
%- {{{ bibliography:

\bibliography{clean}

%- }}}

\end{document}